\documentclass[preprint,12pt]{elsarticle}

\usepackage{graphicx}  
\usepackage{microtype} 
\usepackage{amsmath} 
\usepackage{amssymb}

\usepackage[colorlinks=false,hidelinks]{hyperref}
\usepackage{color}
\usepackage{epstopdf}

\usepackage{natbib}

\newcommand{\beq}{\begin{equation}}  
\newcommand{\eeq}{\end{equation}}    
\newcommand{\beqa}{\begin{eqnarray}} 
\newcommand{\eeqa}{\end{eqnarray}}   

\biboptions{semicolon,square}

\newcounter{bla}

\def\etal{{\it et al.\/}}
\def\ie{{\it i.e.\/}}
\def\eg{{\it e.g.\/}}

\def\1o2{\textstyle {\frac{1}{2}}}

\def\betab{\mbox{\boldmath $\beta$}}
\def\d{{\rm d}}

\def\dotprod{\!\cdot\!}

\def\me{{\rm m}_{\rm e}}

\def\req#1{(\ref{#1})}

\journal{Nuclear Instruments and Methods B}

\begin{document}

\begin{frontmatter}


\title{Electromagnetic interaction models for Monte Carlo simulation of
protons and alpha particles}

\author[a]{Francesc Salvat\corref{author}}
\author[a]{Carlos Heredia}

\cortext[author] {Corresponding author.\\\textit{E-mail address:}
francesc.salvat@ub.edu}

\address[a]{
Facultat de F\'{\i}sica (FQA and ICC), Universitat de Barcelona,
Diagonal 645, 08028 Barcelona, Catalonia, Spain
}

\begin{abstract}
\noindent
Electromagnetic interactions of protons and alpha particles are modeled
	in a form that is suitable for Monte Carlo simulation of the transport
	of charged particles. The differential cross section (DCS) for elastic
	collisions with neutral atoms is expressed as the product of the DCS
	for collisions with the bare nucleus and a correction factor that
	accounts for the screening of the nuclear charge by the atomic
	electrons. The screening factor is obtained as the ratio of the DCS
	for scattering of the projectile by an atom with a point nucleus and
	the parameterized Dirac-Hartree-Fock-Slater (DHFS) electron density,
	calculated from the eikonal approximation, and the Rutherford DCS for
	collisions with the bare point nucleus. Inelastic collisions, which
	cause electronic excitations of the material, are described by means
	of the plane-wave Born approximation, with an empirical simple model
	of the generalized oscillator strength (GOS) that combines several
	extended oscillators with resonance energies and strengths determined
	from the atomic configurations and from the empirical mean excitation
	energy of the material. The contributions from inner subshells are
	renormalized to agree with realistic ionization cross sections
	calculated numerically from the DHFS self-consistent model of atoms by
	means of the plane-wave Born approximation. The resulting DCS allows
	analytical random sampling of individual hard inelastic interactions.
\end{abstract}

\begin{keyword}
Collisions of protons and alphas; Monte Carlo transport of charged
particles; class-II simulation of charged particles.

\end{keyword}

\end{frontmatter}

\section{Introduction \label{sec.1}}

Monte Carlo simulation of the transport of fast charged particles in
matter is
difficult because of the large number of interactions undergone by these
particles before being brought to rest \cite{Berger1963, Jenkins1988}.
This difficulty can be solved by using two alternative strategies: 1)
conventional condensed simulation, or class-I simulation, which consists
of splitting each particle trajectory into a number of steps of definite
length and making use of multiple scattering theories
\cite{GoudsmitSaunderson1940, GoudsmitSaunderson1940b, Landau1944,
Lewis1950} for describing the cumulative effect of the multiple
interactions that occur along each step, and 2) mixed, or class-II
simulation, where hard interactions involving energy transfers or
angular deflections larger than predefined cutoff values are simulated
individually, and soft interactions are described collectively by means
of a multiple-scattering approach \cite{FernandezVarea1993, Salvat2019,
Asai2021}. Class-II schemes are superior because hard interactions are
treated exactly by random sampling from the corresponding restricted
differential cross sections (DCSs), although they require knowledge of the
various DCSs and accurate sampling methods for hard interactions must be
implemented in the simulation code. In the present article we describe
realistic DCSs for elastic and inelastic electromagnetic interactions of
protons and alpha particles with matter, together with algorithms for
the restricted random sampling of hard interactions. The proposed
simulation strategies are applicable to other charged particles
heavier than the electron.

For the sake of generality, the theoretical interaction models are
formulated for the general case of projectile particles with mass $M_1$,
assumed to be larger than the electron mass $\me$, and charge $Z_1
e$, where $e$ denotes the elementary charge. The considered
interactions are elastic collisions with atoms (\ie, interactions that
do not cause excitations of the material) and inelastic interactions,
which result in electronic excitations of the medium. These
interactions are essentially electromagnetic and can be described quite
reliably from first-principles calculations or from appropriate models.

A simulation program transports particles in the laboratory (L) frame,
where the material is at rest and the projectile moves with kinetic
energy $E$ before the interaction. In order to cover the range of
kinetic energies of interest in applications, we shall use relativistic
collision kinematics. For simplicity, we consider that the $z$ axis of
the reference frame is parallel to the linear momentum of the
projectile, which is given by
\beq
{\bf p} = c^{-1} \sqrt{E (E + 2 M_1 c^2)}
\, \hat{\bf z},
\label{a.1}\eeq
where $c$ is the speed of light in vacuum and $M_1$ is the projectile
rest mass,
\beq
M_1 = \left\{ \begin{array}{ll}
m_{\rm p}=1836.15 \; \me \; \; \; & \mbox{for protons,} \\ [1mm]
m_{\rm a}=7294.30\; \me & \mbox{for alphas.}
\end{array} \right.
\label{a.2}\eeq
The rest energy of the electron is $\me c^2 = 511.00$ keV.
The total energy of the projectile is
\beq
{\cal W} = E + M_1 c^2 = \sqrt{M_1^2 c^4 + c^2 p^2}.
\label{a.3}\eeq
We recall the general relations
\beq
p = \beta \gamma \, M_1 c \qquad \mbox{and} \qquad
E=(\gamma-1) M_1 c^2\, ,
\label{a.4}\eeq
where
\beq
\beta= \frac{v}{c} = \frac{\sqrt{E(E+2M_1 c^2)}}{E+M_1 c^2}
\label{a.5}\eeq
is the speed of the particle in units of $c$ and
\beq
\gamma = \sqrt{\frac{1}{1-\beta^2}} = \frac{E+M_1 c^2}{M_1 c^2}
\label{a.6}\eeq
is the particle's total energy in units of its rest energy. The
present article describes the essential physics involved in the
calculation of the DCS and general aspects of the sampling algorithms;
details and specific formulas are given in a document available as
supplementary material.


\section{Elastic collisions}

Let us consider elastic collisions of the projectile with neutral atoms.
These collisions involve a certain transfer of kinetic energy to the
target atom, which manifests as the recoil of the latter. The recoil
of the target atom is easily accounted for by sampling the collisions in
the center-of-mass (CM) frame, which moves relative to the L frame with
velocity
\beq
{\bf v}_{\rm CM} = \betab_{\rm CM} c = \frac{c^2 {\bf p}}{E +
M_1 c^2 + M_{\rm A} c^2},
\label{a.7}\eeq
where $M_{\rm A}$ is the mass of the atom. That is,
\beq
\beta_{\rm CM} =  \frac{v_{\rm CM}}{c} =
\frac{c p}{E + M_1 c^2 + M_{\rm A} c^2} =
\frac{\sqrt{E (E + 2 M_1 c^2)}}{E + M_1 c^2 + M_{\rm A} c^2}\, .
\label{a.8}\eeq

A neutral atom of the element of atomic number $Z$ consists of the
atomic nucleus and $Z$ bound electrons in their ground state. The atomic
nucleus is a system of $Z$ protons and $N$ neutrons, bound together by
the nuclear forces. The total number of nucleons, $A \equiv Z+N$, is
called the {\it mass number}. The atomic mass of the isotope $^AZ$ is
estimated by means a mass formula \cite{RoyerGautier2006} (see the
supplementary document) that approximates the experimental atomic masses
of naturally occurring isotopes \cite{Coursey2015} with a relative
accuracy better than about $10^{-4}$, which is sufficient for the
present purposes.

The calculated cross sections for each element are obtained as an
average over those of the naturally occurring isotopes, weighted by
their respective natural abundances \cite{Coursey2015}. Consistently,
in the simulations we consider that the mass of a target atom is the
average atomic mass of the element \cite{Wang2012}
\beq
M_{\rm A} = \frac{A_{\rm w}}{\rm g/mol} \, {\rm u} \, ,
\label{a.9}\eeq
where $A_{\rm w}$ is the molar mass of the element, and  ${\rm u} =
m(^{12}{\rm C})/12$ is the atomic mass unit. This simplification permits
reducing the required information for each element (and projectile kind)
to a single cross section table, irrespective of the number of isotopes
of that element.

In the CM frame the linear
momenta of the projectile and the atom before the collision are,
respectively, ${\bf p}'_{\rm i} = {\bf p}'_0$ and ${\bf p}'_{\rm Ai} =
-{\bf p}'_0$, with
\beq
{\bf p}'_0 = \frac{M_{\rm A} c^2}{\sqrt{(M_1 c^2
+ M_{\rm A} c^2)^2 + 2 M_{\rm A} c^2 \; E}} \; {\bf p}.
\label{a.10}\eeq
Notice that linear momenta in the CM frame are denoted by primes. After
the elastic collision, in CM the projectile moves with momentum $p'_{\rm
f}=p'_0$ in a direction defined by the polar scattering angle $\theta$
and the azimuthal scattering angle $\phi$, and the target atom recoils
with equal momentum $p'_{\rm Af}=p'_0$ in the opposite direction. The
final energies and directions of the projectile and the atom in the L
frame are obtained by means of a Lorentz boost with velocity $-{\bf
v}_{\rm CM}$.  Thus, elastic collisions are completely determined by the
differential cross section (DCS) per unit solid angle, $\d \sigma/\d
\Omega$, in the CM frame.

We follow the approach described by Salvat and Quesada
\cite{SalvatQuesada2020} (see also Ref.\ \cite{Salvat2022b}), \ie, we
assume that the interaction potential in the CM frame is central, since
this is a prerequisite for applying the partial-wave expansion method to
compute the DCS in the CM frame. Our approach can be qualified as
semi-relativistic, because we are using strict relativistic kinematics
but we do not account for the breaking of the central symmetry of the
interaction when passing from the L to the CM frame.


\subsection{Interaction potential}
\label{sec2.1}

The interaction potential between a charged projectile and the target atom is
expressed as
\beq
V_{\rm scr}(r) = V_{\rm nuc}(r) \,\Phi(r) \, ,
\label{a.11}\eeq
where $r$ is the distance between the projectile and the center of mass
of the atom, $V_{\rm nuc}(r)$ is the interaction energy of the
projectile and the bare atomic nucleus, and
$\Phi(r)$ is the screening function, which accounts for the shielding of
the nuclear charge by the atomic electrons. If the nucleus is
represented as a point structureless charged particle, the nuclear potential
reduces to the Coulomb potential
\beq
V_{\rm nuc}(r) = \frac{Z_1 Z e^2}{r} \equiv V_{\rm C}(r) \, ,
\label{a.12}\eeq
where $Z_1e$ the projectile charge ($Z_1=1$ for protons, $=2$ for
alphas). To facilitate calculations, we use
approximate screening functions having the analytical form
\beq
\Phi(r) = \sum_{i=1}^3 A_i \exp(- a_i r)
\quad \mbox{with} \quad \sum_{i=1}^3 A_i = 1,
\label{a.13}\eeq
with the parameters given by \cite{Salvat1987} for elements with atomic
numbers $Z=1$ to 92, which were determined by fitting the
self-consistent Dirac-Hartree-Fock-Slater (DHFS) atomic potential of
neutral free atoms. Parameters for heavy elements with $Z=93-99$
obtained from the same kind of fit were added more recently. The
advantage of using the representation \req{a.13} of the screening
function is that a good part of the calculation of the DCS for atoms
with point nuclei can be performed analytically \cite{Salvat2022b}.
It is worth noticing that the screened atomic
potential vanishes for radial distances $r$ much larger than the ``atomic
radius'',
\beq
R_{\rm at} \simeq Z^{-1/3} a_0,
\label{a.14}\eeq
where $a_0 = \hbar^2/(\me e^2) = 5.292\times 10^{-9}$ cm is the Bohr
radius.

The interaction energy of the projectile with a bare nucleus of the
isotope $^AZ$ having atomic number $Z$ and mass number $A$ can be
described by a phenomenological complex optical-model potential
\beq
V_{\rm nuc} (r) = V_{\rm opt}(r) + {\rm i} W_{\rm opt}(r),
\label{a.15}\eeq
where the first term is a real potential that reduces to the Coulomb
potential at large radii, and the second term, ${\rm i} W_{\rm nuc}(r)$,
is an absorptive (negative) imaginary potential which accounts for the
loss of projectile particles from the elastic channel caused by
inelastic interactions with the target nucleus. Except for the Coulomb tail, the nuclear potential is of finite-range, it
vanishes when the distance $r$ from the projectile to the nucleus is
larger than about twice the ``nuclear radius'',
\beq
R_{\rm nuc} \sim 1.2 \, A^{1/3} \; \mbox{fm} \, .
\label{a.16}\eeq

Parameterizations of optical-model potentials have been proposed by
various authors. In the calculations for protons (and neutrons) we use
the parameterization of the nuclear global optical-model potential given
by Koning and Delaroche \cite{KoningDelaroche2003}, which is valid for
projectiles with kinetic energies $E$ between 1 keV and about 200 MeV
and nuclei with $24 \le A \le 209$. Owing to the lack of more accurate
approximations, because the potential values vary smoothly with $A$, $Z$
and $E$, we use those parameters for all isotopes with $A>6$ and for
energies up to 300 MeV, for higher energies the potential parameters at
$E=300$ MeV are employed. For protons having $E < 35$ MeV colliding with
target isotopes of mass number $A$ such that $6 < A < 24$ ($Z<12$), we
use the optical-model potential of Watson {\it et al.\/}
\cite{Watson1969}, which is applicable to energies from 10 MeV to 50
MeV; for projectile protons with energies higher than 35 MeV, the
potential of Koning and Delaroche is adopted because it yields DCSs in
better agreement with available experimental information. For alpha
particles, the adopted parameterization of the nuclear potential is the
one proposed by Su and Han \cite{SuHan2015}, which is valid for nuclides
with $20 \le A \le 209$ and projectiles with kinetic energies up to 386
MeV, although we use it for any nucleus. For alphas with higher
energies, we use the parameter values at $E=386$ MeV.

In principle, given the interaction potential, the collision DCS can be
calculated by the method of partial waves
\cite{SalvatFernandezVarea2019}.  As pointed out by Salvat and Quesada
\cite{SalvatQuesada2020}, in the energy range of interest for transport
calculations, the de Broglie wavelength, $\lambda_{\rm dB} = h/p'_0$, of
the projectile is much smaller than the atomic radius $R_{\rm at}$ and,
consequently, the numerical solution of the radial wave equation to
determine the phase-shifts and the DCS is very difficult. In addition,
the partial-wave series converge extremely slowly, requiring the
calculation of a large number ($ \gtrsim 100,000$) of phase-shifts.
Since approximate calculation methods are available for the case of
screened Coulomb potentials (\ie, corresponding to atoms with a point
nucleus), we first calculate the DCS for elastic collisions with bare
nuclei and introduce the effect of electronic screening as a correction
factor to the nuclear DCS.


\subsection{Elastic collisions with bare nuclei}

The scattering of nucleons and alpha particles by nuclei can be described by
using the partial-wave expansion method in the CM frame. The underlying
physical picture is that of a stationary process represented by a
distorted plane wave, \ie, by an exact solution of the time-independent
relativistic Schr\"{o}dinger equation for the potential $V_{\rm
nuc}(r)$,
\beq
\left( - \frac{\hbar^2}{2 \mu_{\rm r}} \, \nabla^2
+ V_{\rm nuc}(r) \right) \psi({\bf r})
= \frac{p'^2_0}{2\mu_{\rm r}} \, \psi({\bf r})
\label{a.17}\eeq
with the relativistic reduced mass
\beq
\mu_{\rm r} = c^{-1} \, \frac{\sqrt{M_1^2 c^2 + p'^2_0} \,
\sqrt{M_{\rm A}^2 c^2 + p'^2_0} }
{\sqrt{M_1^2 c^2 + p'^2_0} + \sqrt{M_{\rm A}^2 c^2 + p'^2_0} },
\label{a.18}\eeq
which asymptotically behaves as a plane wave with an outgoing spherical
wave. Owing to the assumed spherical symmetry of the target nucleus, the
angular distribution of scattered projectiles is axially symmetric about
the direction of incidence, \ie, independent of the azimuthal scattering
angle in both the CM and L frames.

In the case of scattering of spin-unpolarized protons (and neutrons),
the optical-model potential contains
spin-orbit terms, and the wave function is a two-component spinor. The
DCS per unit solid angle in CM is \cite{SalvatFernandezVarea2019}
\beq
\frac{\d \sigma_{\rm nuc}}{\d \Omega} =
\left| f(\theta) \right|^2 + \left| g(\theta) \right|^2.
\label{a.19}\eeq
where the functions $f(\theta)$ and $g(\theta)$ are, respectively,
the direct and spin-flip scattering
amplitudes. They are evaluated from their partial-wave expansions,
\begin{subequations}
\label{a.20}
\beqa
f(\theta) &=&
\frac{1}{2 {\rm i} k} \sum_{\ell}
\left[\rule{0mm}{4mm}
(\ell+1) \left( S_{\ell +} - 1
\rule{0mm}{3.5mm}\right)
+ \ell \left( S_{\ell -} - 1
\rule{0mm}{3.5mm}\right)\rule{0mm}{4mm}
\right] \, P_\ell(\cos\theta)
\label{a.20a}\eeqa
and
\beq
g(\theta) = \frac{1}{2 {\rm i} k}
\sum_{\ell}  \left( S_{\ell +} - S_{\ell -}
\rule{0mm}{3.5mm}\right) \, P_\ell^1(\cos\theta'),
\label{a.20b}\eeq
\end{subequations}
where $P_\ell(\cos\theta')$ and $P_\ell^1(\cos\theta')$
are Legendre polynomials and associated Legendre functions of the first
kind \cite{Olver2010}, respectively, and
\beq
S_{\ell a} = \exp(2 {\rm i} \delta_{\ell a}),
\label{a.21}\eeq
are the $S$-matrix elements. The quantities $\delta_{\ell a}$, with
$a={\rm sign}[2(j-\ell)]$, are the phase-shifts, which depend on the
total and orbital angular momenta of the projectile, $j$ and $\ell$,
respectively. Inelastic interactions with the
nucleus cause a loss of projectile particles from the elastic channel.
The reaction cross section, $\sigma_{\rm react}$, (\ie, the total cross
section for inelastic interactions) is given by
\beq
\sigma_{\rm react} = \frac{\pi}{k^2} \sum_\ell \left\{ (\ell +1)
\left[  1 - |S_{\ell,+}|^2 \rule{0mm}{3.5mm}\right]
+ \ell
\left[  1 - |S_{\ell, -}|^2 \rule{0mm}{3.5mm}\right]
\rule{0mm}{4.5mm}\right\}.
\label{a.22}\eeq
The quantities $T_{\ell a} = 1 - |S_{\ell a}|^2$, the so-called
transmission coefficients, measure the fraction of flux that is lost
from each partial wave.

Since alpha particles have zero spin,
the wave function of these particles is a
scalar. The DCS for elastic collisions of alpha particles with bare
nuclei in the CM frame is given by
\beq
\frac{\d \sigma_{\rm nuc}}{\d \Omega} =
\left| f(\theta) \right|^2
\label{a.23}\eeq
with the scattering amplitude
\beqa
f(\theta) &=&
\frac{1}{2 {\rm i} k} \sum_{\ell}
(2\ell+1) \left( S_{\ell} - 1
\rule{0mm}{3.5mm}\right) P_\ell(\cos\theta) \, ,
\label{a.24}\eeqa
where \cite{SalvatFernandezVarea2019}
\beq
S_{\ell} = \exp(2 {\rm i} \delta_{\ell}).
\label{a.25}\eeq
The reaction cross section for inelastic interactions of alpha particles
with the nucleus is
\beq
\sigma_{\rm react} = \frac{\pi}{k^2} \sum_\ell \left( 2\ell +1 \right)
\left[  1 - |S_{\ell}|^2 \rule{0mm}{3.5mm}\right].
\label{a.26}\eeq

The phase shifts $\delta_{\ell a}$ and $\delta_\ell$ are calculated by
using the Fortran subroutine package {\sc radial} of Salvat and
Fern\'{a}ndez-Varea \cite{SalvatFernandezVarea2019}, which implements a
robust power series solution method that effectively avoids truncation
errors and yields highly accurate radial functions and phase shifts. The
calculations for protons and alpha particles with kinetic energies up to
about 1 GeV are doable because their de Broglie wavelengths are
comparable to the range of the potential (excluding the Coulomb tail,
which determines the kind of ``external'' radial function), $\sim R_{\rm
nuc}$.  It is worth noticing that global optical-model potentials were
adjusted to yield reaction cross sections in agreement with measurements
and, as a consequence, the calculated values of the reaction cross
section and of the DCS are equally reliable.

It is well known that optical-model potentials are not very reliable for
light target nuclei. For collisions of protons with light isotopes
having $A \le 6$ we use the empirical parameterization of the nuclear
DCS described by Galyuzov and Kozov \cite{GalyuzovKozov2021}, which
approximates the available experimental data in an energy range wider
than the one needed for transport calculations. For these light
isotopes, the reaction cross section is estimated from the empirical
formula given by Prael and Chadwick \cite{PraelChadwick1997}.


\subsection{Electronic screening}

Let us consider elastic collisions of the projectile and a
target atom of the element of atomic number $Z$, assuming that the
atomic nucleus can be regarded as a point particle. The corresponding
interaction potential takes the form of a screened Coulomb potential,
\beq
V_{\rm scr}(r) = \frac{Z_1 Z e^2}{r} \, \Phi(r)
= \frac{Z_1 Z e^2}{r} \, \sum_{i=1}^3 A_i \exp(- a_i r),
\label{a.27}\eeq
where we have introduced the analytical screening function \req{a.13}.
The DCS can then be calculated from the wave equation \cite{Salvat2022b}
\beq
\left( - \frac{\hbar^2}{2 \mu_{\rm r}} \, \nabla^2
+ V_{\rm scr}(r) \right) \psi({\bf r})
= \frac{p'^2_0}{2\mu_{\rm r}} \, \psi({\bf r}) \, .
\label{a.28}\eeq

The DCS for collisions of charged particles with a bare point nucleus,
described by the unscreened Coulomb potential $V_{\rm C}(r)$, Eq.\
\req{a.12}, can be obtained from the exact solution of the wave equation
\req{a.28} \cite{Joachain1975} for spinless particles. It is given by
the relativistic Rutherford formula,
\beq
\frac{\d \sigma_{\rm R}}{\d \Omega} = \frac{ \left( 2 \mu_{\rm r} Z_1 Z e^2
\right)^2}{(\hbar q')^4}\, ,
\label{a.29}\eeq
where
\beq
\hbar q' = \left| {\bf p}'_{\rm i} - {\bf p}'_{\rm f} \right| = 2 p'_0
\sin(\theta/2)
\label{a.30}\eeq
is the momentum transfer.

As indicated above, the smallness of the proton wavelength makes the
partial-wave calculation of the DCS for scattering by the screened
Coulomb potential unfeasible. A practical approach adopted in Refs.
\cite{Salvat2013, SalvatQuesada2020} is to use DCSs calculated with the
eikonal approximation \cite{Moliere1947, Schiff1968, Wallace1971}, in
which the phase of the scattered wave is obtained from a semi-classical
approximation to the scattering wave function under the assumption of
small angular deflections of the projectile.

The DCS for scattering by a screened Coulomb potential resulting from
the eikonal approximation is \cite{Salvat2022b}
\beq
\frac{\d \sigma_{\rm scr}}{\d \Omega} =
\left| f_{\rm eik} (\theta) \right|^2.
\label{a.31}\eeq
The function
\beq
f_{\rm eik}(\theta) = - {\rm i} \, k
\int_0^\infty J_0(q'b) \left\{
\exp [{\rm i} \chi(b)]-1\right\} b \, \d b
\label{a.32}\eeq
is the eikonal scattering amplitude at the polar scattering angle
$\theta$ for a particle of mass $\mu_{\rm r}$ and momentum $p'_0 = \hbar
k$.  $J_0(x)$ is the Bessel function of the first kind and zeroth order,
and $\chi(b)$ is the eikonal phase for projectiles incident with impact
parameter $b$. For the analytical potential \req{a.27}, the eikonal
phase takes the form \cite{ZeitlerOlsen1967, Salvat2022b}
\beq
\chi(b) = - \frac{2\mu_{\rm r} Z e^2 }{\hbar^2 k} \sum_i A_i \left\{
K_0(a_i b) - \frac{\mu_{\rm r} Z e^2}{\hbar^2 k^2} \sum_j A_j a_j
K_0[(a_i+a_j)b] \right\},
\label{a.33}\eeq
where $K_0(x)$ is the modified Bessel function of the second kind and
zeroth order. The eikonal scattering amplitude can thus be evaluated by
means of a single quadrature. Because the effect of screening decreases
when the scattering angle increases (\ie, when the classical impact
parameter $b$ decreases), the DCS calculated from the eikonal
approximation, Eq.\ \req{a.31}, tends to the Rutherford DCS at large
angles.

Although the eikonal approximation is expected to be valid for
scattering angles up to about $(k R_{\rm at})^{-1}$ \cite{Moliere1947},
numerical calculations indicate that the approximation yields fairly
accurate DCSs, practically coincident with those obtained from
classical-trajectory calculations up to much larger angles, of the order
of
\beq
\theta_{\rm eik} =
\min \left\{ \frac{200}{kR_{\rm at}}, 0.1 \pi \right\}\, .
\label{a.34} \eeq
For still larger angles the calculation loses validity and presents
numerical instabilities. Following Salvat \cite{Salvat2013}, the DCS for
angles larger than $\theta_{\rm eik}$ is approximated by the expression
\beq
\frac{\d \sigma_{\rm scr}}{\d \Omega} =
\left(\frac{2 \mu_{\rm r} Z e^2}{\hbar^2}\right)^2
\frac{1}{\left[ A + B q'^{2/3} + C q'^{4/3} +
q'^2\right]^2}\, ,
\label{a.35}\eeq
with the coefficients $A$, $B$ and $C$ obtained by matching the
calculated numerical values of the eikonal DCS and its first and second
derivatives at $\theta=\theta_{\rm eik}$. The ratio of the calculated
DCS to the Rutherford DCS,
\beq
F_{\rm scr}(\theta) = \frac{\d \sigma_{\rm scr}}{\d \Omega}
\left( \frac{\d \sigma_{\rm R}}{\d \Omega} \right)^{-1},
\label{a.36}\eeq
measures the effect of screening; it approximates unity at large
angles (see Ref.\ \cite{SalvatQuesada2020}).


\subsection{Elastic-scattering database}
\label{sec2.3}

Considering that 1) the effect of screening is limited to
small angles (large impact parameters), and 2) the DCS for scattering by
the bare finite nucleus differs from the Rutherford DCS only at large angles
(small impact parameters), it follows that screening and nuclear effects
do not interfere. Hence, the CM DCS for collisions of protons and alphas
with neutral atoms can be evaluated as \cite{SalvatQuesada2020}
\beq
\frac{\d \sigma_{\rm el}}{\d \Omega} =
F_{\rm scr}(\theta) \, \frac{\d \sigma_{\rm nuc}}{\d \Omega} \, .
\label{a.37}\eeq
The total elastic cross section is finite and given by
\beq
\sigma_{\rm el} = \int \frac{\d \sigma_{\rm el}}{\d \Omega} \,
\d \Omega
= 2 \pi \int_{-1}^1 \frac{\d \sigma_{\rm el}}{\d \Omega} \, \d
(\cos\theta).
\label{a.38}\eeq

For simulation purposes, it is convenient to consider the DCS as a
function of the angular deflection of the projectile, measured by the
quantity
\beq
\mu \equiv \sin^2(\theta/2) = \frac{1-\cos\theta}{2},
\label{a.39}\eeq
which takes values between 0 (forward scattering) and 1 (backward
scattering). Notice that
\beq
\frac{\d \sigma_{\rm el}}{\d \mu} = \frac{\d \sigma_{\rm el}}{\d \Omega}
\, \frac{2 \pi \, \d (\cos\theta)}{\d \mu} = 4\pi \, \frac{\d
\sigma_{\rm el}}{\d \Omega}\, .
\label{a.40}\eeq
and
\beq
\sigma_{\rm el} = \int_0^1 \frac{\d \sigma_{\rm el}}{\d \mu} \, \d \mu.
\label{a.41}\eeq
We can also write
\beq
\frac{\d \sigma_{\rm el}}{\d \mu} = \sigma_{\rm el} \, p(\mu),
\label{a.42}\eeq
where $p(\mu)$ is the normalized probability density function of $\mu$
in a single collision.

A Fortran program named {\sc panel} has been written to calculate
differential and integrated cross sections for elastic collisions of
protons and alphas (and neutrons) with neutral atoms. This program computes
cross sections for elastic collisions of a projectile particle with a
given isotope $^ZA$ for the kinetic energies of the projectile specified
by the user. Alternatively, it can produce a complete database of DCSs
and integrated cross sections for collisions of projectiles of a given
kind, with laboratory kinetic energies covering the range from 100 keV
to 1 GeV for each element from hydrogen ($Z=1$) to einsteinium ($Z=99$).
As indicated above, the atomic DCSs in the database are obtained as the
average over naturally occurring isotopes of each element.

The database grid of energies is logarithmic, with 35 points per decade.
For each energy the program calculates the DCS in CM, Eq.\ \req{a.37},
for a grid of 1000 polar angles $\theta$. In order to reduce the size
of the database, and also to improve the accuracy of interpolation in
energy, the DCS is tabulated as a function of the variable
\beq
t \equiv 4 (cp'_0)^2 \sin^2(\theta/2) =  4 (cp'_0)^2 \, \mu \, ,
\label{a.43}\eeq
$c^2$ times the square of the momentum transfer in CM. The original
table is ``cleaned'', by removing points in regions where the DCS varies
smoothly, to define a reduced grid that allows accurate natural cubic
spline interpolation in $t$. The DCS interpolated in this way is
estimated to be accurate to four or more digits. For each projectile
energy, the database includes the values of the total elastic cross
section, Eq.\ \req{a.41}, the reaction cross section obtained from Eq.\
\req{a.22} or \req{a.26}, the first transport cross section (or momentum
transfer cross section),
\beqa
\sigma_{{\rm el},1} &\equiv& \int (1-\cos\theta) \,
\frac{\d \sigma_{\rm el}}{\d \Omega} \, \d \Omega
= \int_0^1 2\mu \,\frac{\d \sigma_{\rm el}}{\d \mu} \, \d \mu
\nonumber \\ [2mm]
&=& 2 \sigma_{\rm el} \,  \int_0^1 \mu \,p(\mu) \, \d \mu =
 2 \sigma_{\rm el} \, \langle \mu \rangle \, ,
\label{a.44}\eeqa
and the second transport cross section
\beqa
\sigma_{{\rm el},2} &\equiv& \int \frac{3}{2} \,
\left(1-\cos^2\theta\right) \,
\frac{\d \sigma_{\rm el}}{\d \Omega} \, \d \Omega
\nonumber \\ [2mm]
&=& 6 \sigma_{\rm el} \,  \int_0^1 (\mu-\mu^2) \,p(\mu) \, \d \mu
= 6 \sigma_{\rm el} \left( \langle \mu \rangle
- \langle \mu^2 \rangle \right) ,
\label{a.45}\eeqa
where $\langle \mu^n \rangle$ denotes the $n$-th moment of the angular
deflection in a single collision. The values of these integrated cross
sections serve to assess the accuracy of the DCS interpolation scheme
adopted in the simulation. We recall that the total elastic cross
section and the reaction cross section have the same values in the CM
and L frames.

\begin{figure}[p!] \begin{center}
\includegraphics*[width=8.5cm]{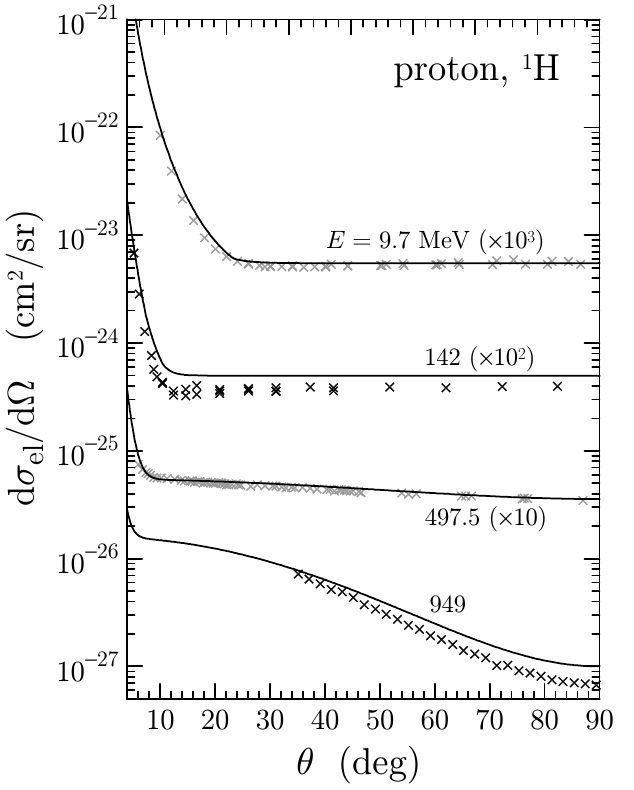}
\caption{
Elastic DCS in the CM frame for collisions of protons with neutral atoms
	of hydrogen, $^1$H. The solid curves are predictions from the
	empirical parameterization of Galyuzov and Kozov
	\cite{GalyuzovKozov2021}. Crosses represent experimental data from
	various authors, as given in the EXFOR database \cite{EXFOR2014}
	(data downloaded from \url{https://www-nds.iaea.org/exfor/} in
	September 2023.).
\label{fig1}}
\end{center} \end{figure}

Figure \ref{fig1} compares results from the empirical formulas of
Galyuzov and Kozov \cite{GalyuzovKozov2021} with experimental data from
various authors, which have been taken from the Experimental Nuclear
Reaction Data (EXFOR) Database of the IAEA \cite{EXFOR2014}
(\url{https://www-nds.iaea.org/exfor/}). The displayed theoretical
curves were obtained by assuming that the projectile and the target atom
are indistinguishable, \ie, the plotted DCS describes collisions where
the projectile is deflected at an angle $\theta$ together with
collisions in which the recoiling target atom moves in directions with
polar angle $\theta$ (or, equivalently, where the projectile emerges in
directions with polar angle $\pi-\theta$). Notice that, as both the
projectile and the recoiling target are followed by the simulation
program, the DCSs in the database are those for the scattered projectile
only, which are defined for $\theta$ between 0 and $\pi$.

\begin{figure}[p!] \begin{center}
\includegraphics*[width=8.5cm]{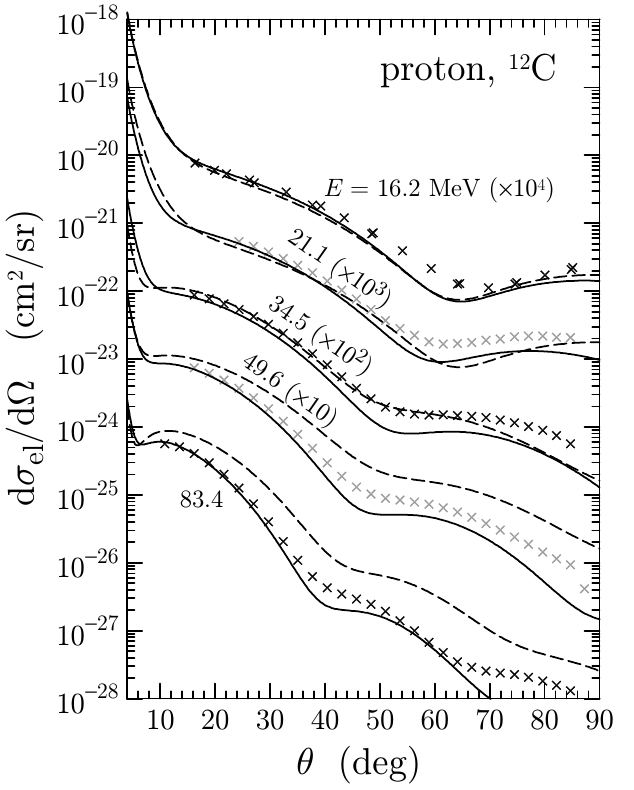}
\caption{
Elastic DCS in the CM frame for collisions of protons with neutral atoms
	of carbon, $^{12}$C.  The solid and dashed curves represent results
	from partial-wave calculations with the nuclear optical-model
	potentials of Koning and Delaroche \cite{KoningDelaroche2003} and of
	Watson {\it et al.\/} \cite{Watson1969}, respectively. Crosses represent
	experimental data from various authors, as given in the EXFOR
	database \cite{EXFOR2014}
	(data downloaded from \url{https://www-nds.iaea.org/exfor/} in
	September 2023.).
\label{fig2}}
\end{center} \end{figure}

As indicated above, collisions of protons with nuclei of light isotopes are
described by means of the optical-model potential of Watson \etal\
\cite{Watson1969} for protons with kinetic energies up to 35 MeV. For
higher energies, the potential of Koning and Delaroche is adopted
\cite{KoningDelaroche2003}. The change of model potential at 35 MeV
is motivated by the comparison of results from both potentials with
experimental data, as illustrated in Fig.\ \ref{fig2}.

\begin{figure}[p!] \begin{center}
\includegraphics*[width=8.5cm]{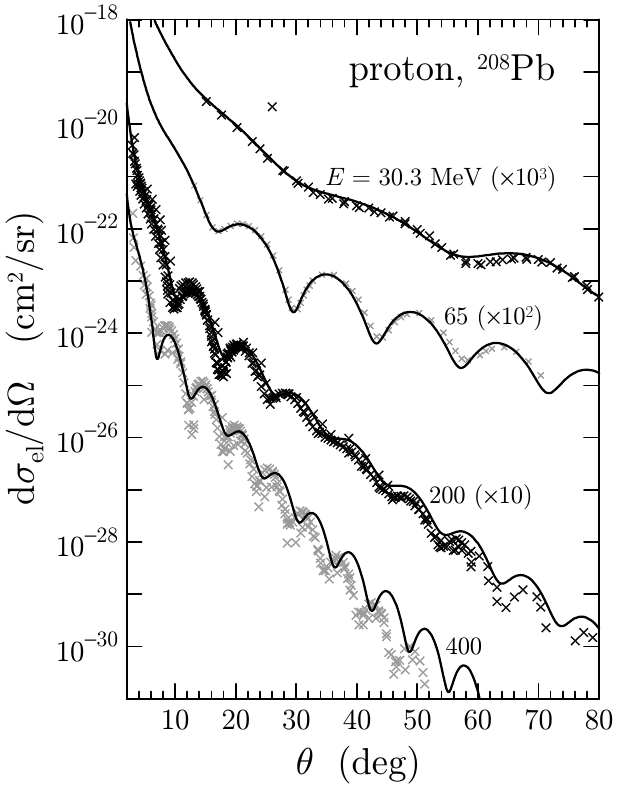}
\caption{
	Elastic DCS in the CM frame for collisions of protons with neutral
	atoms of lead, $^{208}$Pb. The solid curves represent results from
	partial-wave calculations with the global optical-model potential of
	Koning and Delaroche \cite{KoningDelaroche2003}. Crosses represent
	experimental data from various authors, as given in the EXFOR database
	\cite{EXFOR2014} (data downloaded from
	\url{https://www-nds.iaea.org/exfor/} in September 2023.).
\label{fig3}}
\end{center} \end{figure}

The global potential of Koning and Delaroche \cite{KoningDelaroche2003}
is expected to give a quite reliable description of elastic collisions
of protons with isotopes having $A>24$ (which correspond to natural
elements with $Z>11$). This is illustrated in Fig.\ \ref{fig3} for
collisions of protons with atoms of the isotope $^{208}$Pb. Figure 4
compares DCSs of alpha particles with nickel atoms, $^{62}$Ni, with the
nuclear DCS calculated from the optical-model potential of Su and Han
\cite{SuHan2015}, which is expected to provide quite realistic DCSs for
collisions of alphas with any target atom with $A\ge 20$. It is worth
noticing that more reliable theoretical cross sections could be obtained
by using local optical-model potentials (specific of each isotope)
rather than the global potential models adopted here. A partial
justification of the present approach for transport simulations is that
collisions of charged particles much heavier than the electron are
preferentially at small angles, where the DCS is mostly determined by
the screened Coulomb potential of the nucleus; the details of the
nuclear potential affect the DCS only for collisions with intermediate
and large scattering angles, which occur with very small probabilities.

\begin{figure}[p!] \begin{center}
\includegraphics*[width=8.5cm]{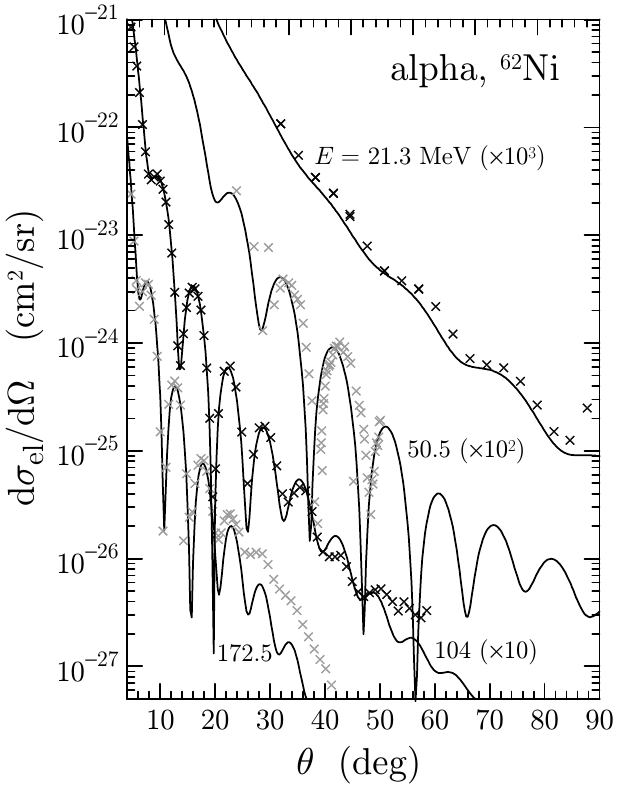}
\caption{
	Elastic DCS in the CM frame for collisions of alpha particles with neutral
	atoms of nickel, $^{62}$Ni. The solid curves represent results from
	partial-wave calculations with the global optical-model potential of
	Su and Han \cite{SuHan2015}. Other details as in Fig. \ref{fig3}.
\label{fig4}}
\end{center} \end{figure}


\subsection{Simulation of elastic collisions}
\label{sec2.4}

Let us assume that the projectile is moving with kinetic energy $E$ in a
compound medium whose molecules consist of $n_i$ atoms of the element
with atomic number $Z_i$ ($i=1, \ldots, N$). The molecular elastic DCS
is obtained from the additivity approximation, \ie, as the sum
of DCSs of the various atoms in a molecule,
\beq
\frac{\d \sigma_{\rm el}}{\d \mu} = \sum_{i=1}^N n_i \,
\frac{\d \sigma_{\rm el}(Z_i)}{\d \mu}
\label{a.46}\eeq
where $\d \sigma_{\rm el}(Z_i)/\d \mu$ denotes the DCS for collisions with
the element of atomic number $Z_i$. The total elastic  molecular cross
sections are expressed similarly,
\beq
\sigma_{\rm el} = \sum_{i=1}^N n_i \,
\sigma_{\rm el}(Z_i)\, ,
\label{a.47}\eeq
and the ratios $p_i= \sigma_{\rm el}(Z_i) / \sigma_{\rm el}$ define the
probabilities of colliding with the various atoms of the molecule. In
accordance with the additivity approximation, we disregard aggregation
effects and, consequently, the atoms in the molecule are assumed to
react as if they were free and at rest.

We consider the detailed simulation of elastic collisions of the
projectile with an atom of the element of atomic number $Z$. The kinematics of
these collisions is completely determined by the polar scattering angle
$\theta$ in CM. In the CM frame, after an elastic collision the
magnitudes of the linear momenta of the projectile and the target atom
are the same as before the collision, and the scattering angles
$\theta$, $\phi$ determine the directions of motion of the two
particles. As mentioned above, the final kinetic energy $E_{\rm f}$ and
the polar scattering angle $\theta_1$ of the projectile in the L frame are
obtained by applying a Lorentz boost with velocity $-{\bf v}_{\rm CM}$.
The final energy of the projectile in L is
\beq
E_{\rm f} = E - W
\label{a.48}\eeq
with the energy loss $W$ given by
\beq
W = W_{\rm max} \, \frac{1-\cos\theta}{2}
= W_{\rm max} \, \mu \, ,
\label{a.49}\eeq
where
\beq
W_{\rm max} = \frac{2M_{\rm A} c^2 \, E(E+2 M_1 c^2)}
{\left( M_1 c^2 + M_{\rm A} c^2 \right)^2 + 2 M_{\rm A} c^2 E}
\label{a.50}\eeq
is the maximum energy loss in a collision, which occurs when $\theta=\pi$.
The polar angle $\theta_1$ of the final direction of the projectile in L
is given by
\beq
\cos\theta_1 = \frac{\tau + \cos\theta}{
\sqrt{(\tau + \cos\theta)^2 + \gamma_{\rm CM}^{-2} \sin^2
\theta}},
\label{a.51}\eeq
with
\beq
\gamma_{\rm CM} \equiv \sqrt{\frac{1}{1-\beta^2_{\rm CM}}} =
\frac{E+M_1 c^2 + M_{\rm A} c^2}{
\left( M_1 c^2 + M_{\rm A} c^2 \right)^2 + 2 M_{\rm A} c^2 E}
\label{a.52}\eeq
and
\beq
\tau = \frac{v_{\rm CM}}{v'_1} = \sqrt{ \left(
\frac{M_1}{M_{\rm A}} \right)^2 (1-\beta_{\rm CM}^2) +
\beta_{\rm CM}^2}\, ,
\label{a.53}\eeq
where
\beq
v'_1 = \frac{c^2 p'_0}{\sqrt{M_1^2 c^4+c^2 p'^2_0}}
\label{a.54}\eeq
is the speed of the scattered projectile in CM.
Notice that the azimuthal angle of the projectile direction in L is the
same as in the CM frame. After the collision, in the L frame the target
atom recoils with kinetic energy $E_{\rm A}=W$ and direction
in the scattering plane with the polar angle $\theta_{\rm A}$ given by
\beq
\cos\theta_{\rm A} = \frac{1-\cos\theta}{\sqrt{(1-\cos\theta)^2
+\gamma_{\rm CM}^{-2} \sin^2 \theta}} \, .
\label{a.55}\eeq

In class II simulations \cite{Salvat2019, Asai2021} it is necessary to
consider the contribution of soft elastic collisions to the elastic
transport cross sections and to the stopping cross section. The required
quantities are determined by the angular DCS in the L frame and by the
energy-loss DCS associated to elastic collisions. The angular DCS is
expressed in terms of the scattering angles in the L
frame by making use of the inverse of the relation \req{a.51},
\beq
\cos \theta = \frac{-\tau \gamma_{\rm CM}^2\sin^2 \theta_1 \pm
\cos\theta_1
\sqrt{\cos^2 \theta_1+\gamma_{\rm CM}^2 (1-\tau^2) \sin^2 \theta_1}}
{\gamma_{\rm CM}^2\sin^2 \theta_1+\cos^2 \theta_1 }\, .
\label{a.56}\eeq
If $\tau$ is less than, or equal to unity only the plus sign
before the square root has to be considered. For $\tau > 1$, there
are two values of the CM deflection $\theta$, given by Eq.\
\req{a.56}, for each value of $\theta_1$, which correspond to different
final energies of the projectile in L. The DCS in the L frame is given by
\beq
\frac{\d \sigma_{\rm el}}{\d \Omega_1} =
\left| \frac{\d (\cos\theta)}{\d (\cos\theta_1)}
\right| \frac{\d \sigma_{\rm el}}{\d \Omega}\, ,
\label{a.57}\eeq
where the last factor is the DCS in the CM frame. From the relation
\req{a.56}, we obtain (a derivation of this expression is given in the
supplementary document)
\beq
\frac{\d \sigma_{\rm el}}{\d \Omega_1} =
\frac{\gamma_{\rm CM}^2\, \left[\tau \cos\theta_1 \pm \sqrt{\cos^2
\theta_1
+\gamma_{\rm CM}^2(1-\tau^2)\sin^2 \theta_1}\; \right]^2}
{\left( \gamma_{\rm CM}^2\sin^2 \theta_1+\cos^2\theta_1 \right)^2
\sqrt{\cos^2\theta_1+\gamma_{\rm CM}^2 (1-\tau^2)\sin^2 \theta_1}} \,
\frac{\d \sigma_{\rm el}}{\d \Omega}.
\label{a.58}\eeq
If $\tau < 1$ only the plus sign is valid and the scattering angle
$\theta_1$ varies from 0 to $\pi$. When $\tau \ge 1$,
the DCS in L vanishes for angles $\theta_1$ larger than
\beq
\theta_{\rm 1,max} = \arccos \left(
\sqrt{\frac{\gamma_{\rm CM}^{2}(\tau^2-1)}
{1+\gamma_{\rm CM}^{2}(\tau^2-1)}}\right)\, ;
\label{a.59}\eeq
for angles $\theta_1 < \theta_{\rm 1,max}$, Eq.\ \req{a.56} yields two
values of $\theta$ in $(0,\pi)$, the expression on the right-hand side
of Eq.\ \req{a.58} must then be evaluated for these two angles (with the
corresponding plus or minus sign in the numerator), and the resulting
values added up to give the DCS in L.

The energy-loss DCS is
\beq
\frac{\d \sigma_{\rm el}}{\d W} = 2\pi \left| \frac{\d W}{\d
(\cos\theta)}\right|^{-1} \frac{\d \sigma_{\rm el}}{\d \Omega} =
\frac{4\pi}{W_{\rm max}} \,
\frac{\d \sigma_{\rm el}}{\d \Omega} = \frac{1}{W_{\rm max}}
\frac{\d \sigma_{\rm el}}{\d \mu} \, ,
\label{a.60}\eeq
and the so-called nuclear stopping cross section is given by
\beqa
\sigma_{\rm el,st} &=& \int_0^{W_{\rm max}} W \,
\frac{\d \sigma_{\rm el}}{\d W} \, \d W
\nonumber \\ [2mm]
&=& W_{\rm max} \sigma_{\rm el} \int_{0}^1
\mu \, p(\mu) \, \d \mu = W_{\rm max} \sigma_{\rm el} \, \langle \mu
\rangle =
\frac{W_{\rm max}}{2} \, \sigma_{\rm el,1},
\label{a.61}\eeqa
where $\sigma_{\rm el,1}$ is the first transport cross section in the CM
frame, Eq.\ \req{a.44}.

The simulation of elastic collisions is performed by using the same
strategy as in the {\sc penelope} and {\sc penh} codes
\cite{Salvat2013, Salvat2019}. Mean free paths and other
energy-dependent quantities are obtained by log-log linear interpolation
of tables, prepared at the start of the simulation run, with a
logarithmic grid of 200 laboratory kinetic energies $E_i$ that covers
the interval of interest. The angular distribution of scattered
projectiles in CM,
\beq
p(E_i, \mu) = \frac{1}{\sigma(E_i)} \, \frac{\d \sigma(E_i)}{\d
\mu}\, ,
\label{a.62}\eeq
is tabulated at the same grid energies.

The CM scattering angle $\theta$ of a projectile with laboratory
energy $E$ in the interval $(E_i, E_{i+1}]$ is sampled from the
distribution
\begin{subequations}
\label{a.63}
\beq
p(E,\mu) = \pi_i \, p(E_i,\mu) + \pi_{i+1} \, p(E_{i+1},\mu)
\label{a.63a}\eeq
with
\beq
\pi_{i} = \frac{ \ln E_{i+1} - \ln E}{\ln E_{i+1} - \ln E_{i}}
\qquad \mbox{and} \qquad
\pi_{i+1} = \frac{ \ln E - \ln E_i}{\ln E_{i+1} - \ln E_{i}}
\label{a.63b}\eeq
\end{subequations}
which is obtained from the tabulated distributions by linear
interpolation in $\ln E$. The sampling is performed by using the
composition method: \\
1) select the value of the index $k=i$ or $i+1$, with respective point
probabilities $\pi_i$ and $\pi_{i+1}$, and \\
2) sample $\mu$ from the distribution $p(E_k,\mu)$. \\
With this interpolation by weight method, $\mu$ is generated by
sampling from only the distributions at the grid energies $E_i$.  This
sampling is performed by the inverse transform method by using the RITA
(rational interpolation with aliasing) algorithm \cite{GarciaTorano2019,
Salvat2019}. The required sampling tables are prepared by the program at
the start of the simulation run.


\section{Inelastic collisions}

Let us now consider the description and simulation of inelastic
collisions of charged particles, \ie, interactions of the projectile
that result in electronic excitations of the material. The most probable
effect of inelastic collisions is the excitation of weakly bound
(valence or conduction) electrons of the material, which can be
described by means of the relativistic plane-wave Born approximation
(PWBA) \cite{Fano1963, Inokuti1971}. Notice that the wave functions of
weakly bound electrons are strongly affected by the state of aggregation
of the material and, hence, a realistic description of the response of
the material requires the use of empirical information. The interaction
model described here accounts for the dependence on the microscopic
structure of the material by using the empirical value of the mean
excitation energy $I$ \cite{ICRU49}, which determines the stopping
power for high-energy projectiles.

Formally, the adopted model is analogous to the one employed in {\sc
penelope} for inelastic collisions of electrons and positrons, which is slightly
modified to yield a finite stopping power for slow projectiles. We
disregard the fact that the mass of the target is finite and,
consequently, inelastic collisions are described in the laboratory
frame, where the stopping material is at rest. For the sake of
generality, we consider a molecular medium, with $Z_{\rm M}$ electrons
in a molecule.  Its electronic structure is described as a number of
bound electron subshells, each with $f_k$ electrons and binding
(ionization) energy $U_k$, which essentially retain their atomic
properties, and, in the case of conducting materials, a set of $f_{\rm
cb}$ nearly free electrons in the  conduction band, with $U_{\rm cb}=0$.
By construction,
\beq
f_{\rm cb} + \sum_k f_k = Z_{\rm M} \, .
\label{a.64}\eeq

Individual inelastic collisions of a projectile (mass $M_1$ and charge
$Z_1 e$) with kinetic energy $E$ and linear momentum ${\bf p}$ are
conveniently characterized by the energy loss of the projectile,
$W=E-E_{\rm f}$, and the magnitude $q$ of the momentum transfer ${\bf q}
\equiv {\bf p} - {\bf p}_{\rm f}$, where $E_{\rm f}$ and ${\bf p}_{\rm
f}$ are, respectively, the kinetic energy and the linear momentum of the
projectile after the interaction. Notice that
\begin{subequations}
\label{a.65}
\beq
(cp)^{2} = E (E+2 M_1 c^{2}),
\label{a.65a}\eeq
and
\beq
(cp_{\rm f})^{2} = (E-W) (E-W+2 M_1 c^{2}).
\label{a.65b}\eeq
\end{subequations}
To simplify the form of the DCS, it is customary to
introduce the so-called recoil energy, $Q$, which is defined as the
kinetic energy of an electron with momentum equal to the momentum
transfer \cite{Fano1963}, in other words,
\beq
Q(Q+2\me c^2) = (cq)^2 = c^2 \left( p^2 + p_{\rm f}^2 - 2 p p_{\rm f} \cos\theta
\right),
\label{a.66}\eeq
where $\theta= \arccos(\hat{\bf p} \dotprod \hat{\bf p}_{\rm f})$ is the polar
scattering angle. Equivalently,
\beq
Q = \sqrt{(cq)^2+\me^2 c^4} - \me c^2.
\label{a.67}\eeq

The doubly-differential cross section (DDCS), differential in $W$ and
$Q$, can be expressed as (see, \eg, \cite{Fano1963, Salvat2019})
\beq
\frac{\d^2 \sigma_{\rm in}}{\d Q \, \d W}
= {\cal B}
\left( \frac{2\me c^2}{WQ(Q+2\me c^2)}
+ \frac{\beta^2 \, \sin^2 \theta_{\rm r} \, W \, 2\me c^2}{[Q(Q+2\me c^2)
- W^2 ]^2} \right) \frac{\d f(Q,W)}{\d W}\, ,
\label{a.68}\eeq
with
\beq
{\cal B} = \frac{2\pi Z_1^2 e^4}{\me v^2} \, ,
\label{a.69}\eeq
and
\beq
\cos^2 \theta_{\rm r} = \frac{W^2/\beta^2}{Q(Q+2\me c^2)}
\left( 1 + \frac{Q(Q+2\me c^2)-W^2}{2W(E+M_1 c^2)} \right)^2,
\label{a.70}\eeq
where $\d f (Q,W)/\d W$ is the generalized oscillator strength (GOS),
which completely characterizes the response of the material. The first
term in expression \req{a.68} describes excitations caused by the
instantaneous Coulomb interaction; the second term accounts for
excitations induced by the transverse interaction (exchange of virtual
photons). We should mention that the transverse contribution in Eq.\
\req{a.68} results from the approximation of neglecting the differences
between longitudinal and transverse GOSs (see, \eg,
\cite{Fano1963,BoteSalvat2008,Salvat2022a}). These differences are
negligible for small $Q$, which dominate in transverse interactions, as
well as for large $Q$.

For a given energy loss $W$, the allowed values of the recoil
energy lie in the interval $(Q_-,Q_+)$, with endpoints given by Eq.\
\req{a.66} with $\cos\theta=+1$ and $-1$, respectively. In other words,
\beq
Q_{\pm} =
\sqrt{\left[cp \pm cp_{\rm f}
\right]^2+\me^2 c^4} - \me c^2.
\label{a.71}\eeq
When $W \ll E$, the lowest allowed recoil energy can be calculated from
the approximate relation \cite{FernandezVarea2005}
\beq
Q_-(Q_-+2\me c^2) = W^2 / \beta^2.
\label{a.72}\eeq
Note that the curves $Q=Q_-(W)$ and $Q=Q_+(W)$ intersect at $W=E$,
Hence, they define a single continuous function $W=W_{\rm m}(Q)$, which
is defined in the interval $0 \le Q \le Q_+(0)$. By solving the
equations $Q=Q_\pm (W_{\rm m})$, we obtain
\beq
W_{\rm m}(Q) = E + M_1 c^2 - \sqrt{\left[ cp -
\sqrt{Q(Q+2\me c^2)} \right]^2 + M_1^2 c^4}\, ,
\label{a.73}\eeq
which, when $W \ll E$, reduces to
\beq
W_{\rm m}(Q) \simeq \beta \sqrt{Q(Q+2\me c^2)}\, .
\label{a.74}\eeq
It follows that, for given values of $E$ and $Q$ [$<Q_+(0)$], the
only kinematically allowed values of the energy loss are those in the
interval $0 < W < W_{\rm m}(Q)$.

The energy-loss DCS is defined by
\beq
\frac{\d \sigma_{\rm in}}{\d W} \equiv \int_{Q_-}^{Q_+}
\frac{\d^2 \sigma_{\rm in}}{\d W \, \d Q} \, \d Q \, .
\label{a.75}\eeq
The probability distribution function (PDF) of the energy loss in a
single inelastic collision is given by
\beq
p_{\rm in} (W) = \frac{1}{\sigma_{\rm in}} \,
\frac{\d \sigma_{\rm in}}{\d W},
\label{a.76}\eeq
where
\beq
\sigma_{\rm in} = \int_0^{W_{\rm max}} \frac{\d \sigma_{\rm in}}{\d W}
\, \d W
\label{a.77}\eeq
is the total cross section for inelastic interactions.
It is convenient to introduce the quantities
\beq
\sigma_{\rm in}^{(n)} \equiv
\int_{0}^{W_{\rm max}} W^{n} \frac{\d\sigma_{\rm in}}{\d W} \, \d W =
\sigma_{\rm in} \int_{0}^{W_{\rm max}} W^{n} p_{\rm
in}(W) \, \d W = \sigma_{\rm in} \, \langle W^n \rangle,
\label{a.78}\eeq
where $\langle W^n \rangle$ denotes the $n$-th moment of the energy loss
in a single collision (notice that $\sigma_{\rm in}^{(0)} = \sigma_{\rm
in}$). $\sigma_{\rm in}^{(1)}$ and $\sigma_{\rm in}^{(2)}$ are known as
the stopping cross section and the energy-straggling cross section,
respectively.

The mean free path $\lambda_{\rm in}$ for inelastic collisions is
\beq
\lambda_{\rm in}^{-1} = {\cal N} \sigma_{\rm in},
\label{a.79}\eeq
where ${\cal N}$ is the number of molecules per unit volume. The
electronic stopping power $S_{\rm in}$ and the energy straggling
parameter $\Omega_{\rm in}^2$ are defined by
\beq
S_{\rm in} = {\cal N}\sigma^{(1)}_{\rm in} = \frac{\langle W
\rangle}{\lambda_{\rm in}} \, ,
\label{a.80}\eeq
and
\beq
\Omega_{\rm in}^2 = {\cal N} \sigma^{(2)}_{\rm in} =
\frac{\langle W^2 \rangle}{\lambda_{\rm in}}\, ,
\label{a.81}\eeq
respectively. The stopping power gives the average energy loss per unit
path length.  The physical meaning of the straggling parameter is less
direct; the product $\Omega_{\rm in}^2(E) \, \d s$ gives the variance of
the energy distribution of charged projectiles that start moving with
energy $E$ after traveling a (small) distance $\d s$ within the medium.


\subsection{The generalized oscillator strength model \label{secgos}}

Although realistic GOSs may be available for simple systems, given
either by analytical formulas (hydrogenic approximation
\cite{Inokuti1971} and electron gas \cite{Lindhard1954}) or by numerical
tables (obtained, \eg, from DHFS calculations for atoms
\cite{BoteSalvat2008, Salvat2022a}), they are not suited for
general-purpose Monte Carlo simulations, mostly because of the strong
correlations between the variables $W$ and $Q$. To account for these
correlations, we should sample the two quantities from their joint PDF,
\ie, from the DDCS, a process that requires massive memory storage and
accurate interpolations.

\begin{figure}[h!] \begin{center}
\includegraphics*[width=6.5cm]{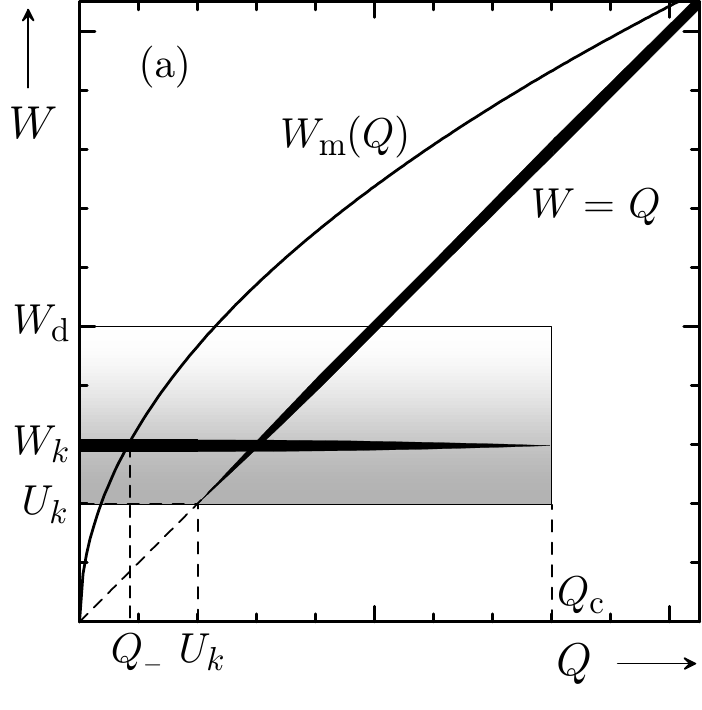} \hfill
\includegraphics*[width=6.5cm]{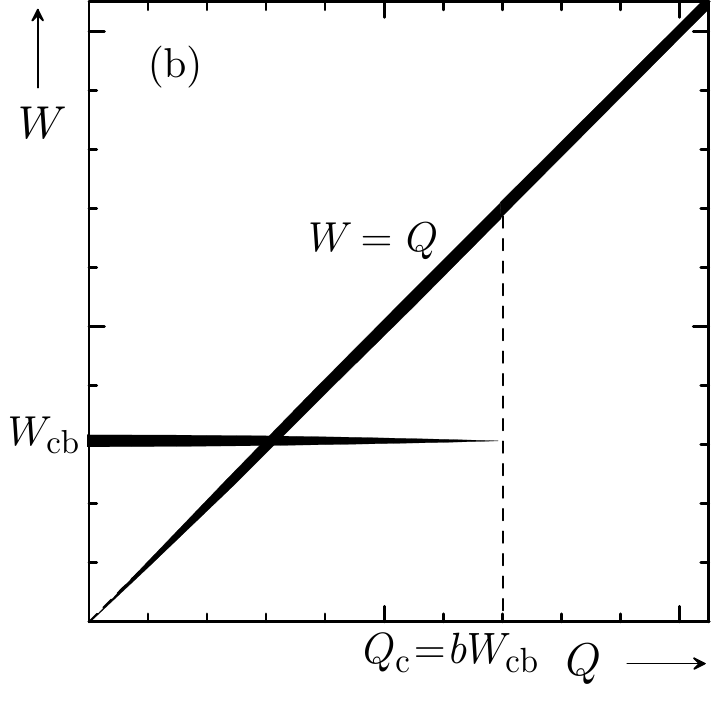}
\caption {Oscillator model for the subshell GOS, represented by the
	solid lines with thickness proportional to the GOS value. The
	continuous curve is the maximum allowed energy loss as a function of
	the recoil energy, $W_{\rm m}(Q)$, Eq.\ \req{a.73} for protons with
	$E=5$ keV.  (a) GOS of a bound subshell with $U_k = 1$ keV. For
	distant interactions the possible recoil energies lie in the interval
	from $Q_-$ to $Q_{\rm c}$, and the energy loss $W$ varies between
	$U_k$ and $W_{\rm d}$, Eq.\ \req{a.114}. (b) Oscillator-GOS model for
	excitations of the conduction band of conductors ($U_{\rm cb}=0$).
\label{fig5}}
\end{center}\end{figure}

Here we use a model of the GOS, adapted from the {\sc
penelope} code \cite{SalvatFernandezVarea1992, Salvat2019}, that
reproduces the most conspicuous features of the GOS, satisfies relevant
sum rules, and provides exact analytical formulas for sampling $W$ and
$Q$ in individual interactions. Excitations of electrons in a subshell
$k$ with binding energy $U_k$ are described as a single ``oscillator''
or one-electron GOS, $F_k(Q,W)$, defined as
\beq
F_k(Q,W) =
\delta(W-W_k) \, g(Q) + \delta(W-Q) \, \left[1 - g(Q) \right],
\label{a.82}\eeq
where $W_k \ge U_k$ and
\begin{subequations}
\label{a.83}
\beq
g(Q)= \left\{
\begin{array}{ll}
1 & \mbox{if $Q < U_k$,} \\ [1mm]
\displaystyle{
1 - \frac{Q^2-U_k^2}{b^2 W_k^2}} \rule{5mm}{0mm}
& \mbox{if $U_k \le Q \le Q_c$,} \\ [1mm]
0 & \mbox{if $Q_c < Q$,}
\end{array}
\right.
\label{a.83a}\eeq
with
\beq
Q_c= \sqrt{b^2 W_k^2+U_k^2}\, .
\label{a.83b}\eeq
\end{subequations}
The quantity $b$ ($>0$) is a free
parameter; a comparison with calculated subshell ionization
cross sections by means of the PWBA with the DHFS potential
\cite{Salvat2022a} (see Fig.\ \ref{fig7} below) indicates that a
value $b\sim 4$ is adequate. Notice that
\beq
\int_0^\infty F_k(Q,W) \, \d W = 1 \qquad \forall Q.
\label{a.84}\eeq
The first term in expression \req{a.82} represents low-$Q$ (distant)
interactions, which are described as a single resonance at the energy
$W_k$. The second term corresponds to large-$Q$ (close) interactions, in
which the target electrons react as if they were free and at rest
($W=Q$); close interactions are allowed only for energy transfers $W$
larger than $U_k$. It is worth noticing that in the case of conductors
the model can be used for describing the GOS of the conduction band
(with $U_{\rm cb} =0$), and the resulting stopping power only vanishes
at $E=0$. Figure \ref{fig5} displays schematically the model GOSs for
inner subshells and for the conduction band.

The molecular GOS is the sum of contributions for
the various electron shells of the atoms in a molecule,
\beq
\frac{\d f(Q,W)}{\d W} = f_{\rm cb} \, F_{\rm cb}(Q,W) + \sum_k f_k
\, F_k(Q,W),
\label{a.85}\eeq
where $f_k$ is the number of electrons in the $k$ subshell.
For bound shells, the resonance energy is defined as
\beq
W_k = \sqrt{(aU_k)^2 + \frac{2}{3} \frac{f_k}{Z_{\rm M}}
\Omega_{\rm p}^2}\, ,
\label{a.86}\eeq
where
\beq
\Omega_{\rm p} = \sqrt{4\pi {\cal N} Z_{\rm M} \hbar^2e^2/\me}
\label{a.87}\eeq
is the plasma energy of a free electron gas with the electron density of
the medium, and $a$ is an adjustable parameter, the so-called
Sternheimer factor. The term $2f_k \Omega_p^2/(3Z_{\rm M})$ in expression
\req{a.86} accounts for the Lorentz-Lorenz correction (the resonance
energies in a condensed medium are larger than those of isolated atoms
or molecules). In the case of conductors, excitations of the conduction
band are represented by a single oscillator with oscillator strength
$f_{\rm cb}$ equal to the number of free electrons per molecule, null
binding energy ($U_{\rm cb}=0$), and resonance energy
\beq
W_{\rm cb} = \sqrt{\frac{f_{\rm cb}}{Z_{\rm M}}} \, \Omega_{\rm p} \, .
\label{a.88}\eeq
Note that $W_{\rm cb}$ is the plasmon excitation energy of a
free-electron gas with the electron density of the conduction band. When
a material is qualified as a conductor, $f_{\rm cb}$ is set equal to the
average lowest negative valence of the elements present ($f_{\rm cb}=0$
for insulators). For free-electron-like materials, such as metallic
aluminum, the value \req{a.88} is close to the energy of volume
plasmons.

The GOS model \req{a.85} satisfies the Bethe sum rule,
\beq
\int_0^\infty \frac{\d f(Q,W)}{\d W} \, \d W
= f_{\rm cb} + \sum_k f_k = Z_{\rm M}
\label{a.89}\eeq
for all $Q$. In the limit $Q \rightarrow 0$ the GOS reduces to the
optical oscillator strength (OOS), which characterizes the optical
properties of the medium, and determines the density effect correction
to the stopping power of high-energy particles. Indeed, the OOS
resulting from our GOS model, with the resonance
energies \req{a.86},
\beq
\frac{\d f(0,W)}{\d W} = f_{\rm cb} \, \delta(W-W_{\rm cb})
+ \sum_k f_k \, \delta(W-W_k) \, ,
\label{a.90}\eeq
coincides with the OOS assumed by Sternheimer \etal\ \cite{Sternheimer1952,
Sternheimer1982} in their calculations of the
density effect correction. The Sternheimer factor $a$ is
fixed by requiring that the GOS model leads to
the empirical value of the mean excitation energy $I$ of the material
\cite{ICRU37},
\beq
Z_{\rm M} \ln I = \int_0^\infty \ln W \,
\frac{\d f(0,W)}{\d W} \, \d W = f_{\rm cb} \ln W_{\rm cb} +
\sum_k f_k \ln W_k \, .
\label{a.91}\eeq
Thus, the GOS is completely determined by the mean excitation energy
$I$, which is the only free parameter of the model. By default the
simulation code uses $I$ values from the ICRU Report 37 \cite{ICRU37}.
Typical values of the Sternheimer factor range between about 2 and 3.
The requirements \req{a.89} and \req{a.91} ensure that the stopping
power of high-energy particles coincides with the values given by the
Bethe formula \cite{SalvatAndreo2023}.


\subsection{Differential and integrated cross sections}

The GOS completely characterizes the response of individual molecules to
inelastic interactions with the projectile (within the PWBA).  The
molecular DDCS can be expressed as
\beq
\frac{\d^2 \sigma_{\rm in}}{\d Q \, \d W} =
\sum_k f_k \frac{\d^2 \sigma_k}{\d Q \, \d W},
\label{a.92}\eeq
where $\d^2 \sigma_k/(\d Q \, \d W)$ is the DDCS for excitations of a
single electron described by the one-electron GOS $F_k(Q,W)$.	Hereafter
the summation over oscillators $k$ includes a term corresponding to the
conduction band, with oscillator strength $f_{\rm cb}$, resonance energy
$W_{\rm cb}$, and ionization energy equal to zero.

The DDCS for collisions with an oscillator is conveniently split into
contributions from close collisions and from distant (resonant)
longitudinal and transverse interactions,
\beq
\frac{\d^2 \sigma_k}{\d Q \, \d W}
= \frac{\d^2 \sigma_k^{\rm c}}{\d Q \, \d W}
+ \frac{\d^2 \sigma_k^{\rm dl}}{\d Q \, \d W}
+\frac{\d^2 \sigma_k^{\rm dt}}{\d Q \, \d W} \, .
\label{a.93}\eeq
The DDCSs for close collisions and for distant longitudinal interactions
are, respectively,
\beq
\frac{\d^2 \sigma_k^{\rm c}}{\d Q \, \d W} =
{\cal B} \, \frac{1}{W^2} \left( 1 - \beta^2 \, \frac{W}{W_{\rm ridge}}
\right) \left[1 - g(Q) \right]
\delta(W-Q) \, \Theta(W_{\rm ridge}-W)
\label{a.94}\eeq
and
\beq
\frac{\d^2 \sigma_k^{\rm dl}}{\d Q \, \d W} =
{\cal B} \, \frac{1}{W} \, \frac{2\me c^2}{Q(Q+2\me c^2)} \,
g(Q) \, \delta(W-W_k) \, \Theta(Q_c-Q).
\label{a.95}\eeq
The quantity $W_{\rm ridge}$ is the maximum energy loss in collisions
of the projectile with free electrons at rest, which is given by
\begin{subequations}
\label{a.96}
\beq
W_{\rm ridge} = 2\me c^2 \beta^2 \gamma^2 \, R
\label{a.96a}\eeq
with
\beq
R \equiv
\left[ 1+\left(\frac{\me}{M_1}\right)^2+ 2 \gamma
\, \frac{\me}{M_1} \right]^{-1}\, .
\label{a.96b}\eeq
\end{subequations}
Notice that, when $M_1=\me$, $W_{\rm ridge}=E$.
For projectiles heavier than the electron ($M_1 \gg \me$) with kinetic
energies much less than their rest energy $M_1c^2$, $R \sim 1$ and
\beq
W_{\rm ridge}\simeq 2\me c^2 \beta^2 \gamma^2  = 2 \me c^2
\left( \gamma^2-1 \right) \, .
\label{a.97}\eeq
The response of molecules in a dense medium is modified by the
dielectric polarization of the material, which modifies the distant transverse
interactions and causes a reduction of the stopping power known as the
{\it density-effect correction}. The DDCS for distant transverse
interactions is approximated as
\beq
\frac{\d^2 \sigma_k^{\rm dt}}{\d Q \, \d W} =
{\cal B} \, \frac{1}{W} \left\{ \ln \left( \frac{1}{1-\beta^2} \right)
- \beta^2 - \delta_{\rm F} \right\}
\delta(W-W_k) \, \delta(Q-Q_-) \, \Theta(Q_c-Q),
\label{a.98}\eeq
where $\delta_{\rm F}$ is the density-effect correction to the stopping
power. It is worth mentioning that this approximate DDCS results from
1) neglecting the angular deflection of the projectile in distant transverse
interactions, which is generally very small, and 2) requiring that it
gives the exact contribution of the distant transverse interactions to
the stopping power for high-energy projectiles, in
accordance with the corrected Bethe formula for the stopping power
\cite{ICRU49}.

The quantity $\delta_ {\rm F}$ is calculated as \cite{InokutiSmith1982,
Salvat2019}
\beq
\delta_{\rm F} \equiv \frac{1}{Z_{\rm M}}
\sum_{k} f_{k} \ln \left( 1+\frac{L^{2}}{W_{k}^{2}} \right)
- \frac{L^{2}}{\Omega_{\rm p}^{2}} \left( 1-\beta^{2} \right),
\label{a.99}\eeq
where $L$ is a real-valued function of $\beta^2$ defined as the positive
root of the equation
\beq
{\cal F}(L) \equiv \frac{1}{Z_{\rm M}} \, \Omega_{\rm p}^{2}
\sum_{k} \frac{f_{k}}{W_{k}^{2}+L^{2}} = 1 - \beta^2.
\label{a.100}\eeq
The function ${\cal F}(L)$ decreases monotonically with $L$, and hence,
the root $L(\beta^{2})$ exists only when $1-\beta^{2}<{\cal F}(0)$;
otherwise $\delta_{\rm F}=0$.
In the high-energy limit ($\beta \rightarrow 1$), the $L$ value
resulting from Eq.\ \req{a.100} is large ($L\gg W_k$) and can be
approximated as $L^2 = \Omega_{\rm p}^2/(1-\beta^2)$. Then, using the
Bethe sum rule \req{a.89} and the relation \req{a.91}, we obtain
\beq
\delta_{\rm F} \simeq \ln \left( \frac{\Omega_{\rm p}^2}{(1-\beta^2)I^2}
\right) - 1, \qquad \mbox{when $\beta \rightarrow 1$}.
\label{a.101}\eeq

The energy-loss DCS for collisions with the $k$-th oscillator, can also
be split into contributions from close, distant longitudinal, and distant
transverse interactions,
\beqa
\frac{\d \sigma_k}{\d W} &=& \int_{Q_-}^{Q_+}
\frac{\d^2 \sigma_k}{\d Q \, \d W} \, \d Q
\nonumber \\ [2mm]
&=& \frac{\d \sigma_k^{\rm c}}{\d W}
+ \frac{\d \sigma_k^{\rm dl}}{\d W}
+ \frac{\d \sigma_k^{\rm dt}}{\d W} \, ,
\label{a.102}\eeqa
where
\beqa
\frac{\d \sigma_k^{\rm c}}{\d W} &=&
\frac{{\cal B}}{W^2} \left( 1 - \beta^2 \, \frac{W}{W_{\rm ridge}}
+ \frac{1-\beta^2}{2 M_1^2 c^4} \, W^2 \right)
\nonumber \\ [2mm]
&& \times \left[1 - g(W) \right]
\Theta(W_{\rm ridge}-W),
\label{a.103}\eeqa
\beqa
\frac{\d \sigma_k^{\rm dl}}{\d W} &=&
\frac{{\cal B}}{W} \left(\int_{Q_-}^{Q_c}
\frac{2\me c^2}{Q(Q+2\me c^2)} \,
g(Q) \, \d Q  \right)
\nonumber \\ [2mm]
&& \times \delta(W-W_k) \, \Theta(Q_c-Q_-),
\label{a.104}\eeqa
and
\beqa
\frac{\d \sigma_k^{\rm dt}}{\d W} &=&
\frac{{\cal B}}{W} \left[ \ln \left( \frac{1}{1-\beta^2} \right)
- \beta^2 - \delta_{\rm F} \right]
\nonumber \\ [2mm]
&& \times \delta(W-W_k) \, \Theta(Q_c-Q_-).
\label{a.105}\eeqa
These energy-loss DCSs, as well as the one-electron cross
sections integrated over an arbitrary interval $(W_1,W_2)$,
\beq
\sigma_k^{(n)} \equiv \int_{W_1}^{W_2} W^n \,
\frac{\d \sigma_k}{\d W} \, \d W ,
\label{a.106}\eeq
can be evaluated analytically (see the supplementary document).

\begin{figure}[h!] \begin{center}
\includegraphics*[width=6.5cm]{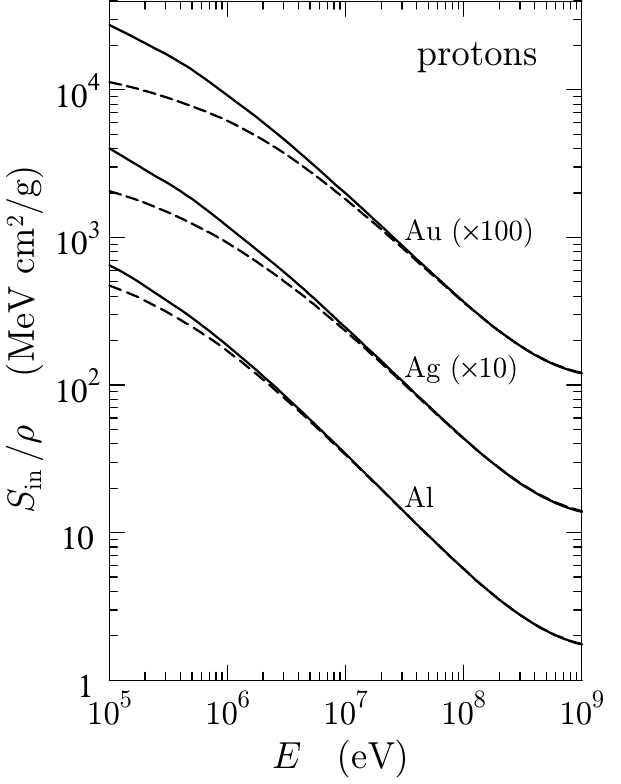}
\includegraphics*[width=6.5cm]{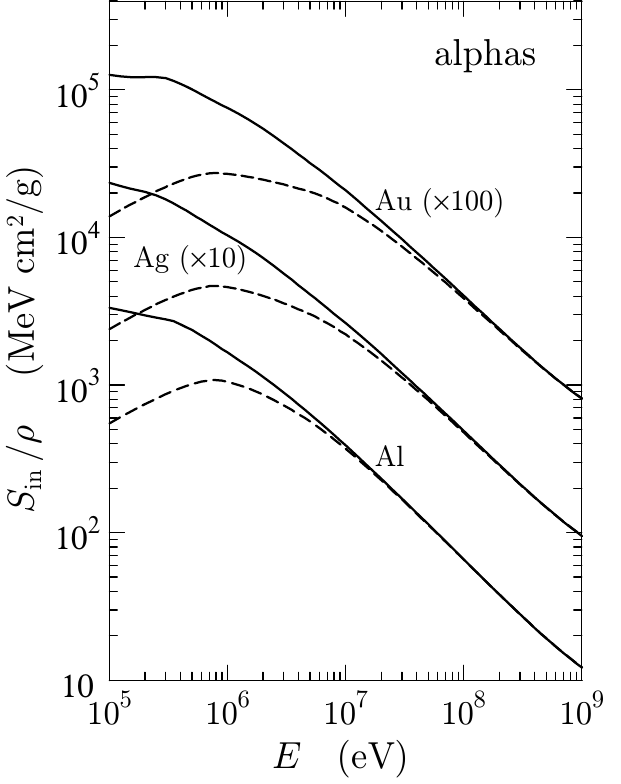}
\caption{
Stopping power of inelastic collisions $S_{\rm in}/\rho$ for protons and
	alpha particles in aluminium, silver ($\times$10) and gold
	($\times$100) as a function of the kinetic energy of the projectile.
	Solid curves are results from the present GOS model. Dashed curves are
	results from the corrected Bethe formula implemented in the program
	{\sc sbethe} \cite{SalvatAndreo2023}.
\label{fig6}}
\end{center} \end{figure}

Evidently, the molecular integrated cross sections for inelastic collisions
are
\beq
\sigma_{\rm in}^{(n)} = \sum_k f_k \, \sigma_{k}^{(n)}.
\label{a.107}\eeq
Figure \ref{fig6} compares the electronic stopping powers of aluminum,
silver, and gold for protons and alpha particles calculated from the
present GOS model with realistic values obtained by means of the program
{\sc sbethe} of Salvat and Andreo \cite{SalvatAndreo2023}, which uses a
corrected Bethe formula. This comparison illustrates our claim that the
stopping power obtained from the GOS model effectively tends to the
realistic value for high-energy projectiles.

\subsubsection{Integrated angular cross sections}

Inelastic collisions cause small deflections of the projectile and
contribute to the directional spreading of particle beams when they
penetrate matter. For simulation purposes, it is convenient to describe
angular deflections by means of the variable $\mu$, Eq.\ \req{a.39},
instead of the polar scattering angle $\theta$. The recoil energy $Q$,
Eq.\ \req{a.66}, can then be expressed as
\beqa
Q(Q+2\me c^{2}) &=& (cp-cp_{\rm f})^{2} + 4 \, cp \, cp_{\rm f} \, \mu .
\nonumber\eeqa
It follows that
\beq
\mu(Q,W) = \frac{Q(Q+2\me c^{2}) - (cp-cp_{\rm f})^{2}}{4 \, cp \,
cp_{\rm f}}\, .
\label{a.108}\eeq
In distant interactions with the $k$-th oscillator, $W=W_k$ and  the
magnitude $p_{{\rm f},k}$ of the linear momentum of the projectile after the
collision,
\beq
(cp_{{\rm f},k})^{2} = (E-W_{k})(E-W_{k}+2\me c^{2}),
\label{a.109}\eeq
is fixed, which implies that $\mu$ is a function of $Q$ only. In close
collisions $Q=W$ and
\beq
\mu(W,W) =
\frac{W(W+2\me c^2) -\left( cp -
\sqrt{(E-W)(E-W+2M_1 c^2)} \right)^2}
{4 \, cp \, \sqrt{(E-W)(E-W +2M_1 c^2)} } \, .
\label{a.110}\eeq

The total angular cross section, the first transport cross section, and the
second transport cross section for inelastic collisions with the $k$-th
oscillator are defined, respectively, as
\begin{subequations} \label{a.111}
\beq
\left[\sigma^{\rm ang}_k \right]^{(0)} =
\int \frac{\d\sigma_{\rm in}}{\d\mu} \, \d\mu \, ,
\label{a.111a}\eeq
\beq
\left[\sigma^{\rm ang}_k \right]^{(1)} =
\int 2\mu \,
\frac{\d\sigma_{\rm in}}{\d\mu} \, \d\mu \, ,
\label{a.111b}\eeq
and
\beq
\left[\sigma^{\rm ang}_k \right]^{(2)} =
\int 6(\mu-\mu^{2})
\frac{\d\sigma_{\rm in}^{\rm (s)}}{\d\mu} \, \d\mu \, ,
\label{a.111c}\eeq
\end{subequations}
where $\d \sigma_{\rm in}/\d \mu$ is the DCS, differential in the
deflection $\mu$. Naturally, both the differential and the integrated
angular cross sections per molecule are the sums of contributions from
the various oscillators,
\beq
\left[\sigma^{\rm ang} \right]^{(n)}
= \sum_k f_k \left[ \sigma^{\rm (ang)}_k \right]^{(n)}.
\label{a.112}\eeq

The contribution of close collisions with the $k$-th oscillator to the
integrated angular cross sections can be calculated in terms of
the energy-loss DCS, while that of distant longitudinal interactions is
conveniently calculated in terms of the DCS differential in the recoil
energy,
\beqa
\frac{\d \sigma_k^{\rm dl}}{\d Q} &=&
\int \frac{\d^2 \sigma_k^{\rm dl}}{\d Q \, \d W} \, \d W
\nonumber \\ [2mm]
&=& \frac{{\cal B}}{W_k} \, \frac{2\me c^2}{Q(Q+2\me c^2)} \,
g(Q) \, \Theta(Q_c-Q)  \, \Theta[Q-Q_-(W_k)]\, .
\label{a.113}\eeqa	
Distant transverse interactions do not contribute to the transport cross
sections because the projectile is not deflected in those interactions.
In the simulation program, the integrals in Eqs.\ \req{a.111} are calculated
numerically (details of this calculation are given in the supplementary
document).

\subsection{Near-threshold distant interactions}

The details of the oscillator GOS model have been tailored to allow
exact random sampling of the energy loss $W$ and the recoil energy $Q$.
In addition, the model can be used for describing interactions with both
bound electrons and conduction electrons. An exact sampling algorithm,
which keeps the correlations between $Q$ and $W$ embodied in the GOS
model, is described in the supplementary document.

Each inelastic interaction with the $k$-th oscillator causes the release
of a secondary electron with kinetic energy $E_{\rm s}=W - U_k$ in the
direction of the momentum transfer, defined by the polar angle
$\theta_{\rm r}$ given by Eq.\ \req{a.70}.

In the case of excitations of a bound subshell, the energy loss
distribution associated with distant interactions is described as a
single resonance (delta function), while the actual distribution is
continuous for energy losses above the ionization threshold. As a
consequence, energy loss spectra simulated from the present GOS model will
show unphysical narrow peaks at energy losses that are multiples of the
resonance energies. To get rid of this kind of artifact, we spread the
resonance line by sampling the energy loss in distant interactions from
the continuous triangular distribution in the interval from $U_k$ to
\beq
W_{\rm d} = 3 W_k  - 2 U_k.
\label{a.114}\eeq
That is, we consider the distribution
\beq
p_{\rm d}(W) = \frac{2}{(W_{\rm d} - U_k)^2}\, (W_{\rm d} - W),
\label{a.115}\eeq
which gives the correct average value, $\left< W \right> = W_k$ (see
Fig.\ \ref{fig5}). Since energy losses larger than $W_{\rm m} (Q_{\rm
c})$ are forbidden, the value of $W_{\rm d}$ should be smaller than
$W_{\rm m} (Q_{\rm c})$. When this is not the case,
we modify the resonance energy $W_k$, and replace it with the value
\beq
W'_k = \left\{ \begin{array}{ll}
U_k & \mbox{if $W_{\rm m} (Q_{\rm c}) \le U_k$,} \\ [2mm]
	[W_{\rm m} (Q_{\rm c}) + 2U_k]/3 \; \; \; & \mbox{if $U_k < W_{\rm m} (Q_{\rm c}) \le 3 W_k  - 2
U_k$,} \\ [2mm]
W_k & \mbox{if $3 W_k  - 2 U_k < W_{\rm m} (Q_{\rm c})$,}
\end{array} \right.
\label{a.116}\eeq
That is, the quantity $W_k$ is replaced with this modified value in all
formulas pertaining to the distant excitations of bound subshells. Also,
to prevent an anomalous increase of the ionization cross section of
bound subshells for projectiles with kinetic energy near the threshold,
we multiply the DCS for distant excitations by the factor
\beq
N^{\rm d,thres.} = \left( \frac{W'_k - U_k}{W_k - U_k} \right)^2,
\label{1.117}\eeq
which reduces to unity when $W_{\rm m} (Q_{\rm c})$ is larger than
$3 W_k  - 2 U_k$.

Thus, the maximum allowed energy loss in distant excitations of bound
subshells, Eq. \req{a.114}, is given by
\beq
W_{\rm d} = 3W'_k - 2U_k,
\label{a.118}\eeq
which never exceeds $W_{\rm m} (Q_{\rm c})$. The energy loss in distant
excitations is sampled from the pdf \req{a.115} by using the sampling
formula
\beq
W = W_{\rm d} - \left( W_{\rm d} - U_k \right) \sqrt{\xi},
\label{a.119}\eeq
where $\xi$ is a random number uniformly distributed in (0,1); this
formula results from the inverse transform method \cite{Salvat2019}. The spread
distribution and the low-energy modification of the resonance energy are
applied only to bound electron subshells. The energy spectrum of distant
interactions with conduction-band electrons is not altered, \ie, the
energy loss in these excitations equals $W_{\rm cb}$ independently of the
energy of the projectile.


\subsection{Ionization of inner subshells and re-normalization}

The GOS model given by Eq.\ \req{a.85} provides a quite realistic
description of the correlations between the energy loss and the
scattering angle in inelastic collisions of charged particles. However,
the subshell total cross section obtained from that GOS model may differ
appreciably from results of experiments and of more accurate calculations.
Inaccuracies in the total cross section for ionization of inner electron
subshells become apparent when we consider the emission of x rays
induced by impact of charged particles: the number of x rays
emitted is proportional to the ionization cross section of the active
subshell.

To provide a more accurate description of the emission of x rays and
Auger electrons, we have calculated a complete database of cross
sections for ionization of inner subshells (K shell, L, M, and N
subshells with binding energy larger than 50 eV) of all the elements
from hydrogen to einsteinium ($Z=1$ to 99), by impact of protons and
alpha particles with energies up to 10 GeV. The calculations were based
on the relativistic PWBA, as formulated by Bote and Salvat
\cite{BoteSalvat2008} (see also \cite{Salvat2022a}) using
longitudinal and transverse GOSs computed with the DHFS potential.
Following Chen \cite{Chen1983} and Chen and Crasemann
\cite{ChenCrasemann1989}, we adopted the perturbed-stationary-state
approximation of Brandt and Lapicki \cite{BrandtLapicki1979}, which
improves the PWBA by accounting for (1) alterations in the binding of
the active electron due to the presence of the projectile near the
nucleus of the target atom, and (2) the deflection of the projectile
path caused by the Coulomb field of the nucleus. Details of these
calculations are described by Salvat \cite{Salvat2021}. Chen and
Crasemann \cite{ChenCrasemann1989} performed similar calculations using
the non-relativistic PWBA, also with GOSs obtained from the DHFS
potential, and published tables of cross sections for ionization by
protons with energies up to 5~MeV. Our results agree closely with
theirs, but extend to much higher energies. In addition, to
approximately account for the density effect, we reduce the cross
sections in the database by a factor equal to the ratio of the cross
sections obtained from the GOS model with and without the density effect
correction, $\delta_{\rm F}$.  Hereafter, the ionization cross section
of our calculated database, with this density-effect correction factor,
will be referred to as ``reference'' ionization cross sections.

\begin{figure}[h!] \begin{center}
\includegraphics*[width=6.5cm]{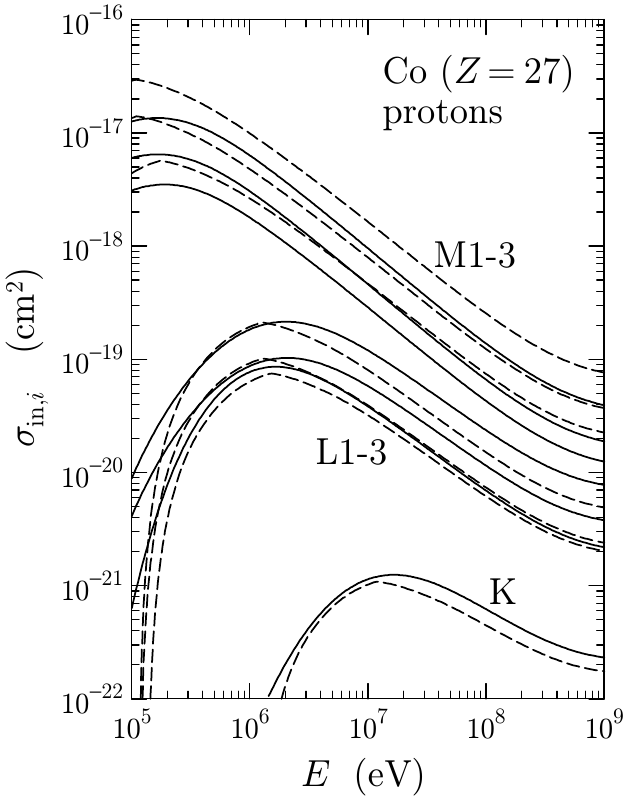}
\includegraphics*[width=6.5cm]{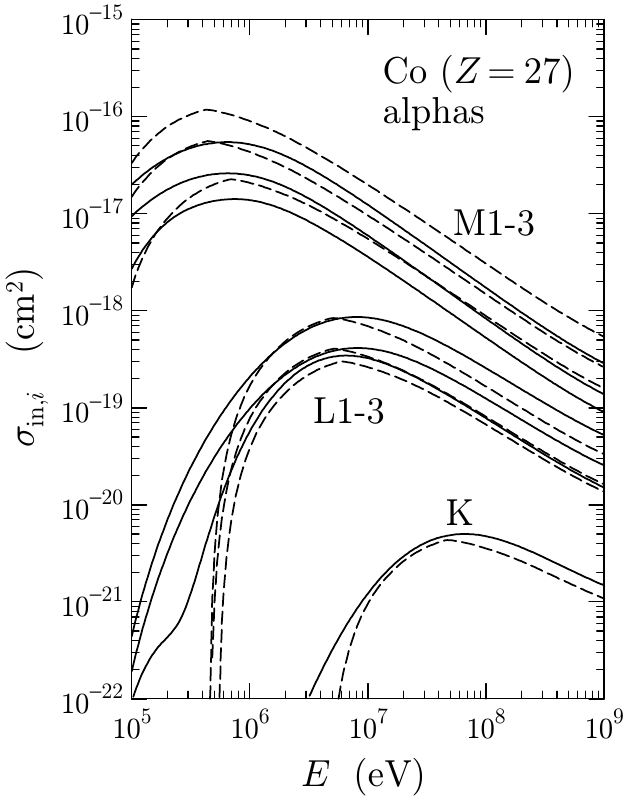}
\caption{
Ionization cross sections of the inner subshells of cobalt atoms by
	impact of protons and alphas, as functions of the kinetic energy of
	the projectile. Solid curves represent the reference ionization cross
	sections obtained from the accurate calculations described in the
	text. Dashed curves are the predictions from the present GOS model
	for solid cobalt.
\label{fig7}}
\end{center} \end{figure}

In our simulation program, the total cross section, $\sigma_{\rm in}$, is
decomposed into contributions from inner and outer electron subshells,
\beq
\sigma_{\rm in} (E) =  \sum_i f_i \, \sigma_{{\rm in,}i} (E)
+ \sum_j f_j \, \sigma_{{\rm in,}j} (E) ,
\label{a.120}\eeq
where the first summation is over inner subshells (\ie, K to N7
subshells with binding energies $U_i$ greater than the cut-off energy
$E_{\rm cut}=50$~eV); the second summation is over outer subshells (\ie,
those with $U_j < E_{\rm cut}$ or with principal quantum number larger
than 4).  Figure \ref{fig7} compares the reference ionization cross
sections of the inner shells of the cobalt atom ($Z=27$) with the
predictions of our GOS model for solid cobalt. The various curves
correspond to the indicated subshells; notice that $\sigma_{{\rm in,}i}$
tends to increase when the binding energy of the active subshell
decreases. As the total cross section and the stopping cross section are
dominated by contributions from outer subshells with relatively small
binding energies, the total cross sections of inner subshells may be
modified, up to a certain extent, and those of the outer subshells may
be re-normalized so that the input stopping power remains unaltered.

The simulation program assumes that hard inelastic collisions with inner
subshells ionize the target atom, and the relaxation of the resulting
vacancies is simulated by the {\sc penelope} routines by using the
transition probabilities given in the Evaluated Atomic Data Library of
Perkins \etal\ \cite{Perkins1991}. To get the correct number of emitted
x rays, the total cross section of each inner shell, $f_i \sigma_{{\rm
in,}i} (E)$, is replaced with the reference cross section $\sigma_{{\rm
in},i}^{\rm
(ref)}(E)$, without altering the details of the PDF of the energy-loss
and scattering angle.  That is, the  ``oscillator strength'' $f_i$ of
the $i$-th inner shell is replaced with
\beq
f'_i =
\frac{\sigma_{{\rm in},i}^{\rm (ref)}(E)}{\sigma_{{\rm in},i} (E)} \, ,
\label{a.121}\eeq
when $\sigma_{{\rm in},i} (E) >0$. It is worth noticing that because of
the neglect of the motion of atomic electrons in close collisions, the
GOS model gives effective ionization thresholds that are higher than
those of the reference cross sections. That is, we may have
$\sigma_{{\rm in,}i}(E)=0$ but $\sigma_{{\rm in},i}^{\rm (ref)}(E) \ne
0$, in which case the projectile particles can ionize the inner shell at
energies lower than the corresponding ionization threshold; under these
circumstances, the energy transfer is set equal to the binding energy of
the subshell, $W=U_i$, and the projectile's trajectory  is not
deflected. Of course, this procedure implies increasing the
inner-subshell contribution to the stopping power in the (small)
quantity $U_i \, \sigma_i^{\rm (ref)}(E)$.

The program reads a table of the stopping power, $S_{\rm in}(E)$, from
the input material-data file, which is considered to be the actual
stopping power of the material. By default, this table is calculated
from the GOS model \req{a.85} as described above. In order to avoid
altering the input stopping power, the total cross sections of outer
subshells, $f_j \, \sigma_j(E)$, are multiplied by an energy-dependent
scaling factor, $N(E)$, the same for all outer subshells, given by
\beq
N(E) = \left[ S_{\rm in}(E) - \sum_i f'_i \, \sigma_{{\rm in,}i}^{(1)}(E)
\right] \left( \sum_j f_j \, \sigma_{{\rm in,}j}^{(1)}(E) \right)^{-1} ,
\label{a.122}\eeq
where $\sigma_{{\rm in,}j}^{(1)} (E)$ is the one-electron stopping cross
section for excitations of the $j$-th outer subshell, Eq.\ \req{a.106}.
Formally, this modification is equivalent to replacing the oscillator
strengths $f_j$ of the outer subshells with $f'_j = N(E) \, f_j$.

As already mentioned, by default the input stopping power is calculated
from the PWBA with the GOS model \req{a.85}. However, the PWBA with the
density-effect correction is valid only for projectiles with relatively
high energies. Departures from the PWBA give rise to the Lindhard-S\o
rensen and Barkas corrections to the Bethe formula
\cite{SalvatAndreo2023}. To account for these departures, the user may
edit the input material-data file and replace the stopping power table
with more reliable values. As reference stopping powers one may use
those generated by the program {\sc sbethe} of Salvat and Andreo
\cite{SalvatAndreo2023}, which are consistent with the recommendations
and values given in the ICRU Report 49 \cite{ICRU49}.


\section{Tracking algorithm}

The interaction models described above permit the formulation of a
class-II tracking scheme \cite{Salvat2019, Asai2021} with a fixed
energy-loss cutoff $W_{\rm cc}$, which is set by the user, and an
energy-dependent cutoff deflection $\mu_{\rm c}$ for elastic collisions
that is defied internally by the program in terms of two user-defined
simulation parameters, $C_1$ and $C_2$. Particle trajectories are
generated by using the random-hinge method
\cite{FernandezVarea1993,Salvat2019}, which operates similarly to
detailed simulations, \ie, the transported particle is moved in straight
``jumps'', and the energy and direction of movement change only through
discrete events (hard interactions and hinges). Here we sketch the
simulation algorithms briefly, additional details can be found in the
manual of the code system {\sc penelope} and in the article by Asai
\etal\ \cite{Asai2021}.

\subsection{Elastic collisions}

\begin{figure}[h!] \begin{center}
\includegraphics*[width=6.5cm]{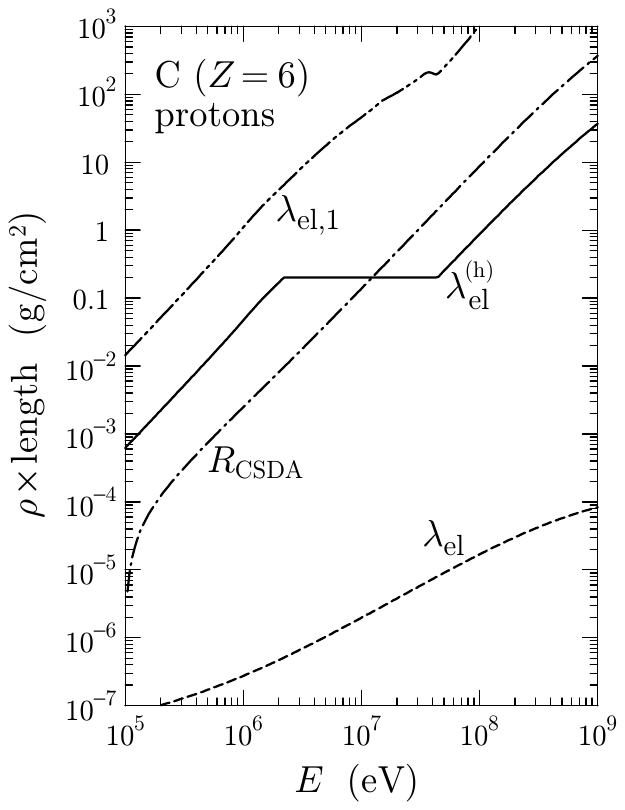}
\includegraphics*[width=6.5cm]{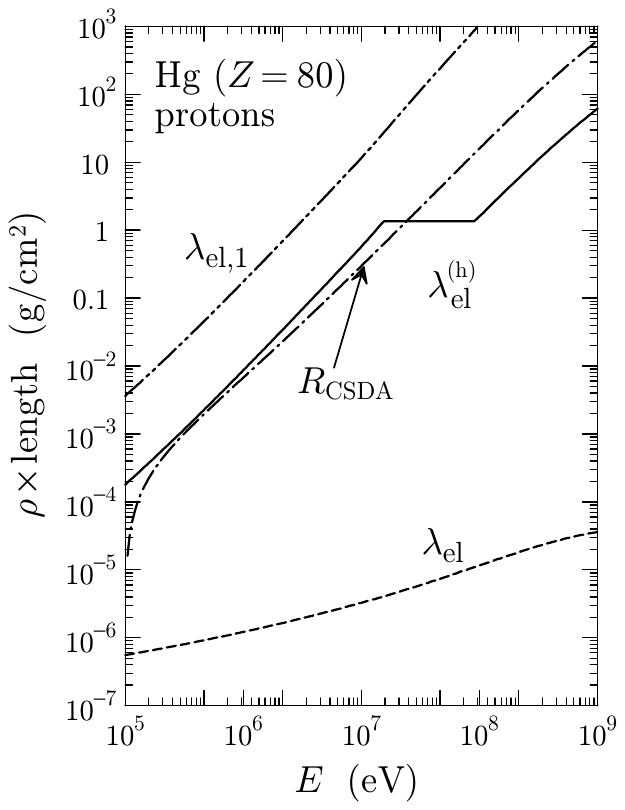}
\caption{
Elastic mean free path $\lambda_{\rm el}$, first transport mean free
path $\lambda_{{\rm el},1}$ and range $R_{\rm CSDA}$ of protons in carbon and
mercury.  The solid curves represent the mean free path between hard
elastic events $\lambda_{\rm el}^{\rm (h)}$ obtained from Eq.\
\req{a.123} with $C_{1}=0.05$ and $C_{2}=0.10$.
\label{fig8}}
\end{center} \end{figure}

In our simulation code the cutoff deflection $\mu_{\rm c}$, which
separates hard and soft elastic collisions, is determined by two
energy-independent user parameters, $C_1$ and $C_2$, which typically
should be given small values, between 0 and 0.2. These two parameters
are used to fix the mean free path between hard elastic events (\ie, the
average step length between consecutive hard elastic collisions), which
is defined as
\beq
\lambda_{\rm el}^{\rm (h)} = {\rm max} \left\{
\lambda_{\rm el}, {\rm min}\left[
	C_1 \lambda_{\rm el,1}, C_2 \, \max \left( R_{\rm CSDA}, \mbox{1 cm}
	\right) \right] \right\},
\label{a.123}\eeq
where $\lambda_{{\rm el},1}=[{\cal N} \sigma_{{\rm el},1}]^{-1}$ is the
first transport mean free path, see Eq.\ \req{a.44}, and
\beq
R_{\rm CSDA} (E) = \int_{E_{\rm abs}}^E \frac{\d E'}{S_{\rm in}(E')}
\label{a.124}\eeq
is the CSDA range calculated from the input electronic stopping power.
The identity
\beq
\lambda_{\rm el}^{\rm (h)} (E) = \left[
{\cal N}  \int_{\mu_{\rm c}}^1 \frac{\d
\sigma_{\rm el} (E)}{\d \mu} \, \d \mu \right]^{-1}\,
\label{a.125}\eeq
then fixes the cutoff $\mu_{\rm c}$ as a function of the energy $E$ of the
projectile, which may be different for the various atoms in a molecule.
The recipe \req{a.123} forces high-energy particles to proceed in steps
of average length $\lambda_{\rm el}^{\rm (h)} =C_2 \, R_{\rm CSDA}$,
while low-energy projectiles have the average step length $\lambda_{\rm
el}^{\rm (h)} = C_1 \lambda_{\rm el,1}$. Figure \ref{fig8} illustrates
the situation for protons in carbon and mercury, as representatives of
low- and high-$Z$ solid materials. The transition between the high- and
low-energy ranges corresponds to the horizontal segment in the plots,
where $\lambda_{\rm el}^{\rm (h)} =C_2$ cm.  Hence, $C_1$ only affects
particles with low energies, while $C_2$ effectively controls the
simulation of high-energy projectiles.

The average angular
deflection of the particle trajectory at the end of a step of length
$\lambda_{\rm el}^{\rm (h)}$ can be evaluated from Lewis'
theory \cite{Lewis1950} which, ignoring energy losses along the step, gives
\beq
1- \langle \cos\theta_{\rm m} \rangle = 1- \exp\left( -
\, \frac{\lambda_{\rm el}^{\rm (h)}}{\lambda_{{\rm el},1}}
\right) \simeq \frac{\lambda_{\rm el}^{\rm (h)}}{\lambda_{{\rm
el},1}} \lesssim C_1.
\label{a.126}\eeq
That is, $C_1$ sets an approximate upper limit for the average angular
deflection (measured in the CM frames) at the end of the step. On the
other hand, $C_2$ limits the average fractional energy loss along the
step. An increase of $C_1$ or $C_2$ leads to increased values of both
the mean free path between hard events, $\lambda_{\rm el}^{\rm (h)}$,
and the cutoff deflection, $\mu_{\rm c}$, in certain energy ranges
\cite{Salvat2019}. Of course, an increase of $\lambda_{\rm el}^{\rm
(h)}$ implies a reduction in the number of hard events along a particle
track with an accompanying reduction of the simulation time.

The angular deflection effect of the soft interactions that occur
between each consecutive pair of hard interactions is determined by the
transport cross sections of orders $\ell=0$ and 1 of the soft
interactions in the L frame. The contributions from elastic collisions
are
\beq
\sigma_{{\rm el},\ell}^{\rm (s)} (E) =
\int_0^{\mu_{{\rm c},1}} \left[ 1 - P_\ell (\cos\theta_1) \right]
\frac{\d \sigma_{\rm el}(E)}{\d \mu_1} \,
\d \mu_1 \, ,
\label{a.127}\eeq
where $\mu_1$ is the angular deflection in the L frame.
It is important to notice that soft inelastic collisions also cause a
small deflection of the projectile. The scattering effect of these
interactions is accounted for by considering their contributions to the
soft transport cross sections,
\beq
\sigma_{{\rm in},\ell}^{\rm (s)} (E) =
\int_0^1 \left[ 1 - P_\ell (\cos\theta) \right] \frac{\d \sigma_{\rm in}(E)}
{\d \mu} \, \d \mu,
\label{a.128}\eeq
where
\beq
\frac{\d \sigma_{\rm in}(E)}{\d \mu} =
\sum_k \frac{\d \sigma_k}{\d \mu} \, \Theta(W_{\rm cc} - W)
\label{a.129}\eeq
is the sum of contributions of all oscillators restricted to energy
losses less than $W_{\rm cc}$.
The combined (elastic plus inelastic) soft scattering process is then
described by the transport mean free paths
\beq
\frac{1}{\lambda_{{\rm comb},\ell}^{\rm (s)} (E)} = {\cal N}
\left[ \sigma_{{\rm el},\ell}^{\rm (s)} (E) +
\sigma_{{\rm in},\ell}^{\rm (s)} (E) \right]
\label{a.130}\eeq
of orders $\ell =1$ and 2.
Assuming that the energy loss is small, the first and second moments of
the angular deflection after a path length $s$, under the sole action of
soft elastic and soft inelastic interactions, are \cite{Lewis1950,
Salvat2019}
\begin{subequations}
\label{a.131}
\beq
\langle\mu_{\rm s} \rangle =
\frac{1}{2} \left[ 1 - \exp(-s/\lambda_{{\rm comb},1}^{\rm (s)}) \right]
\label{a.131a}\eeq
and
\beq
\langle\mu^{2}_{\rm s} \rangle =
\langle\mu_{\rm s}\rangle
- \frac{1}{6} \left[ 1 - \exp(-s/\lambda_{{\rm comb},2}^{\rm (s)})
\right].
\label{a.131b}\eeq
\end{subequations}

In practical simulations the angular deflection $\mu_{\rm s}$ after a
path length $s$ is sampled from an artificial distribution, $P(\mu_{\rm
s})$, which is required to have the same moments,
\beq
\left< \mu_{\rm s}^n \right> =
\int_0^1 \mu_{\rm s}^n \, P(\mu_{\rm s}) \, \d \mu_{\rm s},
\label{a.132}\eeq
of orders $n=1$ and 2 as the real distribution, Eqs.\ \req{a.131}, but is
otherwise arbitrary \cite{Salvat2019, Asai2021}.

\subsection{Inelastic collisions}

As indicated above, the simulation of inelastic collisions is tuned by
the cutoff energy transfer $W_{\rm cc}$ set by the user, which separates
soft and hard interactions. Hard inelastic interactions with energy-loss higher
than $W_{\rm cc}$ are simulated individually from the corresponding
restricted DDCS. To simplify the programming, distant interactions with
an oscillator are considered to be hard only if $U_k \ge W_{\rm cc}$, \ie,
distant excitations of oscillators with $U_k < W_{\rm cc}$ are all soft.
This classification avoids the need of splitting the continuous distribution
\req{a.115}. The sampling of hard interactions is performed exactly by
using the algorithms described in the supplementary document, modified so as to deliver energy
losses larger than $W_{\rm cc}$. Along each trajectory step (to or from
a hard interaction), soft interactions with $W < W_{\rm cc}$ may occur.
The cumulative effect of these soft interactions is described by means
of a multiple scattering approach determined by the {\it restricted}
stopping power,
\beq
S_{\rm in}^{\rm (s)} (E) =
{\cal N} \int_0^{W_{\rm cc}}
W \, \frac{\d \sigma_{\rm in} (E)}{\d W} \, \d W
\label{a.133}\eeq
and the {\it restricted} energy straggling parameter,
\beq
\Omega^{2{\rm (s)}}_{\rm in} (E) =
{\cal N} \int_0^{W_{\rm cc}} W^2 \,
\frac{\d \sigma_{\rm in} (E)}{\d W} \, \d W.
\label{a.134}\eeq
For the sake of numerical consistency, we also include the stopping due
to soft elastic collisions, which accounts for energy transfers
$W=W_{\rm max} \, \mu$ to recoiling target nuclei (nuclear stopping) ,
\beq
S_{\rm el}^{\rm (s)} (Z,E) = {\cal N}
\int_0^{\mu_{\rm c}}
W \, \frac{\d \sigma_{\rm el} (Z,E)}{\d \mu} \, \d \mu \, ,
\label{a.135}\eeq
\beq
\Omega^{2{\rm (s)}}_{\rm el} (Z,E) =
{\cal N} \int_0^{\mu_{\rm c}}
W^2 \, \frac{\d \sigma_{\rm el} (Z,E)}{\d \mu} \, \d \mu \, ,
\label{a.136}\eeq
where both $W_{\rm max}$, Eq.\ \req{a.50}, and $\mu_{\rm c}$, Eq.
\req{a.125},
are specific of each target element. The global stopping power and
energy-straggling parameter of soft interactions are
\beqa
S_{\rm s}(E) &=& S_{\rm in}^{\rm (s)} (E) + S_{\rm el}^{\rm (s)} (E)\, ,
\nonumber \\ [2mm]
\Omega^2_{\rm s}(E_0) &=& \Omega^{2{\rm (s)}}_{\rm in} (Z) +
\Omega^{2{\rm (s)}}_{\rm el} (E) \, .
\label{a.137}\eeqa
A difficulty of class-II algorithms arises from the fact that the energy
of the particle decreases along the step between two consecutive hard
interactions. Because the cutoff energy $W_{\rm cc}$ does not change with
$E$, we can assume that, at least for small
fractional energy losses, the DCSs for soft energy-loss events vary
linearly with $E$. Under this assumption we can calculate the first
moments of the distribution of the energy loss $W_{\rm s}$ of a
particle with initial energy $E_0$ after traveling a path length $s$
under only the influence of soft events \cite{Salvat2019}. The
mean and variance of this distribution are, respectively,
\begin{subequations}
\label{a.138}
\beq
\langle W_{\rm s} \rangle =
S_{\rm s}(E_0) \, s \left\{ 1 - \frac{1}{2}
\left[\frac{\d \ln S_{\rm s}(E)}{\d E}
\right]_{E=E_0} S_{\rm s}(E_0) \, s \right\} \rule{5mm}{0mm}
\label{a.138a}\eeq
and
\beq
{\rm var}(W_{\rm s}) =
\Omega^2_{\rm s}(E_0) \, s \left\{ 1 -
\left[\frac{1}{2} \frac{\d \ln \Omega^2_{\rm s}(E)}{\d E}
+ \frac{\d \ln S_{\rm s}(E)}{\d E} \right]_{E=E_0} S_{\rm s}(E_0) \, s
\right\},
\label{a.138b}\eeq
\end{subequations}
where the factors in curly braces account for the global effect of the
energy dependence of the soft energy-loss DCS, within the linear
approximation.

The energy loss caused by soft events along a trajectory step is sampled
from an artificial pdf with parameters obtained from the stopping cross
section and the energy-straggling cross section for soft interactions
\cite{Salvat2019}. The accumulated angular deflection caused by soft
interactions along a step is sampled from an artificial distribution
with its first and second moments determined by the first and second
transport cross sections restricted to soft interactions. These integral
characteristics of soft interactions are readily obtained from the
expressions given above with the appropriate limits of the integrals.


\section{Concluding comments}

We have presented DCSs for elastic and inelastic collisions of protons
and alpha particles suited for class-II Monte Carlo simulations of the
transport of charged particles in matter. The DCS for elastic collisions
are calculated from realistic nuclear optical-model potentials by using highly
accurate partial-wave methods, and corrected to account for the effect
of screening of the nuclear charge by the atomic electrons. Atomic DCSs
in the CM frame have been calculated for the elements with atomic
numbers 1 to 99; they have been included in an extensive database for
protons, alpha particles (and neutrons) with kinetic energies between
100 keV and 1 GeV.

Inelastic collisions are described by means of the PWBA, in order to
provide a description of electron binding effects and of the correlations
between the energy loss and the deflection angle of the projectile in
inelastic events. The proposed GOS model satisfies the Bethe sum rule,
and partially incorporates the effect of aggregation by using an
empirical value of the mean excitation energy $I$ as a defining
parameter. As a consequence our
DCSs lead to the correct electronic stopping for high energy
projectiles. A simple renormalization of the DCS of inner subshells, to
agree with ionization cross sections calculated with the DHFS
self-consistent potential, ensures that simulations will generate the
correct number of ionizations and the ensuing emission of x rays and
Auger electrons. In addition, a further renormalization of the DCSs of
outer electron subshells permits incorporating more realistic stopping
powers for projectiles with intermediate and low energies.

The proposed interaction models can be used in class-II simulations of
charged-particle transport. They permit the formulation of adequate
sampling algorithms for hard interactions, \ie, elastic collisions with
angular deflections larger than $\mu_{\rm c}$ and inelastic collisions
with energy loss larger then $W_{\rm cc}$, with arbitrary cutoffs. An
exact sampling algorithm for inelastic collisions is described in the
supplementary document. These models and databases have been implemented
in a Fortran simulation code named {\sc penhan} that, in conjunction
with {\sc penelope} \cite{Salvat2019}, simulates the coupled transport
of electrons, positrons, photons, protons, and alpha particles in
matter. A detailed description of {\sc penhan}, which is available from
the authors under request, will be published elsewhere.


\section*{Acknowledgments}
\addcontentsline{toc}{section}{Acknowledgments}

We are thankful to Dr A. A. Galyuzov for providing a Fortran subroutine
with the parameterized DCS for proton-nucleus elastic collisions.
Financial support from the Spanish Ministerio de Ciencia e
Innovaci\'{o}n / Agencia Estatal de Investigaci\'{o}n / European
Regional Development Fund, European Union, (project no.\ PID2021-123879
OB-C22) is gratefully acknowledged.




\begin{center}
\rule[-0.125mm]{1.5cm}{0.25mm}\rule[-0.3mm]{4.5cm}{.6mm}\rule[-0.125mm]{1.5cm}{0.25mm}
\end{center}

\end{document}